\documentclass[longauth]{aa}
\usepackage[utf8]{inputenc}
\usepackage[varg]{txfonts}
\usepackage[breaklinks, colorlinks, citecolor=blue, linkcolor=blue]{hyperref}
\usepackage[usenames, dvipsnames]{color}
\usepackage{amsmath}
\usepackage{graphicx}
\usepackage[english]{babel}
\usepackage{siunitx}
\usepackage{float}
\usepackage{url}
\usepackage{longtable}
\usepackage{fancyhdr}
\usepackage{natbib}
\usepackage[normalem]{ulem}
\usepackage[version=4]{mhchem}

\makeatletter

\def\instrefs#1{{\def\scsep{\def\scsep{,}}\@for\w:=#1\do{\scsep\ref{inst:\w}}}}

\renewcommand{\inst}[1]{\unskip$^{\instrefs{#1}}$}
\newcommand*\samethanks[1][\value{footnote}]{\footnotemark[#1]}

\let\orgautoref\autoref
\renewcommand{\autoref}
        {\def\equationautorefname{Eq.}
         \def\figureautorefname{Fig.}
         \def\sectionautorefname{Sect.}
         \def\subsectionautorefname{Sect.}
         \def\subsubsectionautorefname{Sect.}
         \orgautoref}

\renewcommand*\aa@pageof{, page \thepage{} of \pageref*{LastPage}}

\makeatother

\begin{document}

\title{An ultra-short-period transiting super-Earth\\ orbiting the M3 dwarf TOI-1685}
\titlerunning{Ultra-short period transiting super-Earth}

\author{
P.~Bluhm\inst{lsw}\thanks{Fellow of the International Max Planck Research School for Astronomy and Cosmic Physics at the University of Heidelberg (IMPRS-HD).}
\and E.~Pall\'e\inst{iac,ull} 
\and K.~Molaverdikhani\inst{lsw,mpia}
\and J.~Kemmer\inst{lsw}\samethanks
\and A.\,P.~Hatzes\inst{tls}
\and D.~Kossakowski\inst{mpia}\samethanks
\and S.~Stock\inst{lsw}\samethanks
\and J.\,A.~Caballero\inst{cabesac}
\and J.~Lillo-Box\inst{cabesac}
\and V.\,J.\,S.~B\'ejar\inst{iac,ull}
\and M.\,G.~Soto\inst{qmul}
\and P.\,J.~Amado\inst{iaa}
\and P.~Brown \inst{LCO_A}
\and C.~Cadieux\inst{OMM}
\and R.~Cloutier\inst{harv}\thanks{Banting Fellow}
\and K.\,A.~Collins\inst{harv}
\and K.\,I.~Collins\inst{gmu}
\and M.~Cort\'es-Contreras\inst{cabesac}
\and R.~ Doyon\inst{OMM}
\and S.~Dreizler\inst{iag}
\and N.~Espinoza\inst{stsci}
\and A.~Fukui\inst{depst,ull}
\and E.~Gonz\'alez-\'Alvarez\inst{cabinta}
\and Th.~Henning\inst{mpia}
\and K.~Horne\inst{supa}
\and S.\,V.~Jeffers\inst{mig}
\and J.\,M.~Jenkins\inst{Ames}
\and E.\,L.\,N.~Jensen\inst{dpa}
\and A.~Kaminski\inst{lsw}
\and J.\,F.~Kielkopf\inst{louis}
\and N.~Kusakabe\inst{naoj}
\and M.~K\"urster\inst{mpia}
\and D.~Lafreni\`ere\inst{OMM}
\and R.~Luque\inst{iac,ull}
\and F.~Murgas\inst{iac,ull}
\and D.~Montes\inst{ucm}
\and J.\,C.~Morales\inst{ice,ieec}
\and N.~Narita\inst{ikis,jst,osa,iac}
\and V.\,M.~Passegger\inst{hs,uok}
\and A.~Quirrenbach\inst{lsw}
\and P.~Sch\"ofer\inst{iag}
\and S.~Reffert\inst{lsw}
\and A.~Reiners\inst{iag}
\and I.~Ribas\inst{ice,ieec}
\and G.\,R.~Ricker\inst{kavli}
\and S.~Seager\inst{kavli,MIT1,MIT2}
\and A.~Schweitzer\inst{hs}
\and R.\,P.~Schwarz\inst{Voor}
\and M.~Tamura\inst{osa,naoj,ut}
\and T.~Trifonov\inst{mpia}
\and R.~Vanderspek\inst{kavli}
\and J.~Winn\inst{prince}
\and M.~Zechmeister\inst{iag}
\and M.\,R.~Zapatero~Osorio\inst{cabinta}
}

\institute{
\label{inst:lsw}Landessternwarte, Zentrum f\"ur Astronomie der Universit\"at Heidelberg, K\"onigstuhl 12, 69117 Heidelberg, Germany \\
\email{pbluhm@lsw.uni-heidelberg.de}
\and 
\label{inst:iac}Instituto de Astrof\'isica de Canarias (IAC), E-38200 La Laguna, Tenerife, Spain
\and 
\label{inst:ull}Departamento de Astrof\'isica, Universidad de La Laguna, 38206 La Laguna, Tenerife, Spain
\and 
\label{inst:mpia}Max-Planck-Institut f\"ur Astronomie, K\"onigstuhl 17, 69117 Heidelberg, Germany
\and
\label{inst:tls}Th\"uringer Landessternwarte Tautenburg, Sternwarte 5, 07778 Tautenburg, Germany
\and
\label{inst:cabesac}Centro de Astrobiolog\'ia (CSIC-INTA), ESAC, Camino bajo del castillo s/n, 28692 Villanueva de la Ca\~nada, Madrid, Spain
\and
\label{inst:qmul}School of Physics and Astronomy, Queen Mary University London, 327 Mile End Road, London E1 4NS, UK
\and
\label{inst:iaa}Instituto de Astrof\'isica de Andaluc\'ia (CSIC), Glorieta de la Astronom\'ia s/n, 18008 Granada, Spain
\and
\label{inst:LCO_A}{Department of Physics and Astronomy, Vanderbilt University, 6301 Stevenson Center Ln., Nashville, TN 37235, USA}  
\and
\label{inst:OMM}Institute for Research on Exoplanets (IREx), Universit\'e de Montr\'eal, D\'epartement de Physique, C.P.~6128 Succ. Centre-ville, Montr\'eal, QC H3C~3J7, Canada 
\and
\label{inst:harv}
Center for Astrophysics \textbar Harvard \& Smithsonian, 60 Garden Street, Cambridge, MA 02138, USA
\and
\label{inst:gmu}Department of Physics and Astronomy, George Mason University, 4400 University Drive, Fairfax, VA 22030, USA
\and
\label{inst:iag}Institut f\"ur Astrophysik, Georg-August-Universit\"at, Friedrich-Hund-Platz 1, 37077 G\"ottingen, Germany
\and
\label{inst:stsci}Space Telescope Science Institute, 3700 San Martin Drive, Baltimore, MD 21218, USA
\and
\label{inst:depst}Department of Earth and Planetary Science, Graduate School of Science,  The  University of Tokyo, 7-3-1 Hongo, Bunkyo-ku, Tokyo 113-0033, Japan
\and
\label{inst:cabinta}Centro de Astrobiolog\'ia (CSIC-INTA), Carretera de Ajalvir km 4, 28850 Torrej\'on de Ardoz, Madrid, Spain
\and
\label{inst:supa}SUPA Physics and Astronomy, University of St. Andrews, Fife, KY16 9SS, Scotland, UK
\and
\label{inst:mig}{Max-Planck Institute for Solar System Research
Justus-von-Liebig Weg 3,
37077 Goettingen}
\and
\label{inst:Ames}NASA Ames Research Center, Moffett Field, CA 94035, USA
\and 
\label{inst:dpa}Department of Physics \& Astronomy, Swarthmore College, Swarthmore, PA 19081, USA
\and
\label{inst:louis}Department of Physics and Astronomy, University of Louisville, Louisville, KY 40292, USA
\and
\label{inst:naoj}National Astronomical Observatory of Japan, 2-21-1 Osawa, Mitaka, Tokyo 181-8588, Japan.
\and
\label{inst:ucm}Departamento de F\'{i}sica de la Tierra y Astrof\'{i}sica and IPARCOS-UCM (Instituto de F\'{i}sica de Part\'{i}culas y del Cosmos de la UCM), Facultad de Ciencias F\'{i}sicas, Universidad Complutense de Madrid, 28040 Madrid, Spain
\and
\label{inst:ice}Institut de Ci\`encies de l’Espai (ICE, CSIC), Campus UAB, Can Magrans s/n, 08193 Bellaterra, Barcelona, Spain
\and 
\label{inst:ieec}Institut d’Estudis Espacials de Catalunya (IEEC), 08034 Barcelona, Spain
\and
\label{inst:ikis}Komaba Institute for Science, The University of Tokyo, 3-8-1 Komaba, Meguro, Tokyo 153-8902, Japan
\and
\label{inst:jst}Japan Science and Technology Agency, PRESTO, 3-8-1 Komaba, Meguro, Tokyo 153-8902, Japan
\and
\label{inst:osa}Astrobiology Center, 2-21-1 Osawa, Mitaka, Tokyo 181-8588, Japan
\and
\label{inst:hs}Hamburger Sternwarte, Universit\"at Hamburg, Gojenbergsweg 112, 21029 Hamburg, Germany
\and
\label{inst:uok}Homer L. Dodge Department of Physics and Astronomy, University of Oklahoma, 440 West Brooks Street, Norman, OK 73019, USA
\and
\label{inst:kavli}Kavli Institute for Astrophysics and Space Research, Massachusetts Institute of Technology, Cambridge, MA 02139, USA
\and
\label{inst:MIT1}Department of Earth, Atmospheric and Planetary Sciences, Massachusetts Institute of Technology, Cambridge, MA 02139, USA
\and
\label{inst:MIT2}Department of Aeronautics and Astronautics, Massachusetts Institute of Technology, 77 Massachusetts Avenue, Cambridge, MA 02139, USA
\and
\label{inst:Voor}Patashnick Voorheesville Observatory, Voorheesville, NY 12186, USA
\and
\label{inst:ut}Department of Astronomy, The University of Tokyo, 7-3-1 Hongo, Bunkyo-ku, Tokyo 113-0033, Japan
\and
\label{inst:prince}Princeton University, Princeton, NJ 08544, USA
}

\date{1 March 2021 / Accepted dd Month 2021}

\abstract{Dynamical histories of planetary systems, as well as the atmospheric evolution of highly irradiated planets, can be studied by characterizing the ultra-short-period planet population, which the {\em TESS} mission is particularly well suited to discover.
Here, we report on the follow-up of a transit signal detected in the {\em TESS} sector 19 photometric time series of the M3.0\,V star TOI-1685 (2MASS J04342248+4302148).  
We confirm the planetary nature of the transit signal, which has a period of $P_{\rm b}$~=~$ 0.6691403^{+ 0.0000023}_{- 0.0000021}$\,d, using precise radial velocity measurements taken with the CARMENES spectrograph. 
From the joint photometry and radial velocity analysis, we estimate the following parameters for TOI-1685\,b: a mass of $M_{\rm b}$~=~$3.78 ^{+ 0.63}_{- 0.63}$\,$M_\oplus$, a radius of $R_{\rm b}$~=~$1.70^{+ 0.07}_{- 0.07}$\,$R_\oplus$, which together result in a bulk density of $\rho_{\rm b}$~=~$4.21^{+0.95}_{-0.82}$\,g\,cm$^{-3}$, and an equilibrium temperature of $T_\textnormal{eq}$~=~$1069 ^{+16}_{-16}$\,K. 
TOI-1685\,b is the least dense ultra-short-period planet around an M dwarf known to date. 
TOI-1685\,b is also one of the hottest transiting super-Earth planets with accurate dynamical mass measurements, which makes it a particularly attractive target for thermal emission spectroscopy. 
Additionally, we report with moderate evidence an additional non-transiting planet candidate in the system, TOI-1685\,[c], which has an orbital period of $P_{\rm c}$~=~$9.02^{+ 0.10}_{- 0.12}$\,d.
}
\keywords{planetary systems --
    techniques: photometric --
    techniques: radial velocities --
    stars: individual: TOI-1685 --
    stars: late-type
    }

\maketitle

\section{Introduction}\label{sec:intro}

Currently, over one hundred planets with orbital periods of less than one day are known\footnote{\url{https://exoplanetarchive.ipac.caltech.edu/}, \\ \url{http://exoplanet.eu/}}. These exoplanets, normally referred to as ultra-short-period planets \citep[USPs;][]{2006Natur.443..534S, 2018NewAR..83...37W}, are frequently found around main-sequence stars.
The majority of USPs are small ($R \le$~2--3\,$R_{\oplus}$) and appear to have compositions similar to that of the Earth \citep{2018NewAR..83...37W}. Their origin is still uncertain. 
One possible scenario is that these planets were originally hot Jupiters that experienced a phase of intense erosion due to tidal activity and/or intense stellar irradiation \citep{2013ApJ...775..105O}, while in another scenario the progenitors of USPs were the exposed remnants of so-called mini-Neptunes, which can still harbor external gaseous layers
\citep[]{2016NatCo...711201L, 2017ApJ...842...40L}.
Additional theories propose that these objects might have formed at more separated orbits before migrating to their current locations
\citep{2015MNRAS.448.1729R, 2017ApJ...842...40L} or even formed in situ \citep{2013MNRAS.431.3444C}.
For the moment, a clear picture of the origins of these objects remains elusive \citep{2015MNRAS.446.3676A}, which makes them critical tracers of theories of planet formation and evolution.
Due to their proximity to their host stars, these planets can reach equilibrium temperatures of thousands of kelvins \citep{2011ApJ...741L..30R,2012ApJ...751L..28D, 2013ApJ...774...54S}, which also makes them ideal laboratories for studying atmospheric composition via thermal emission spectroscopy. 

\begin{figure*}
    \centering
    \includegraphics[width=\textwidth]{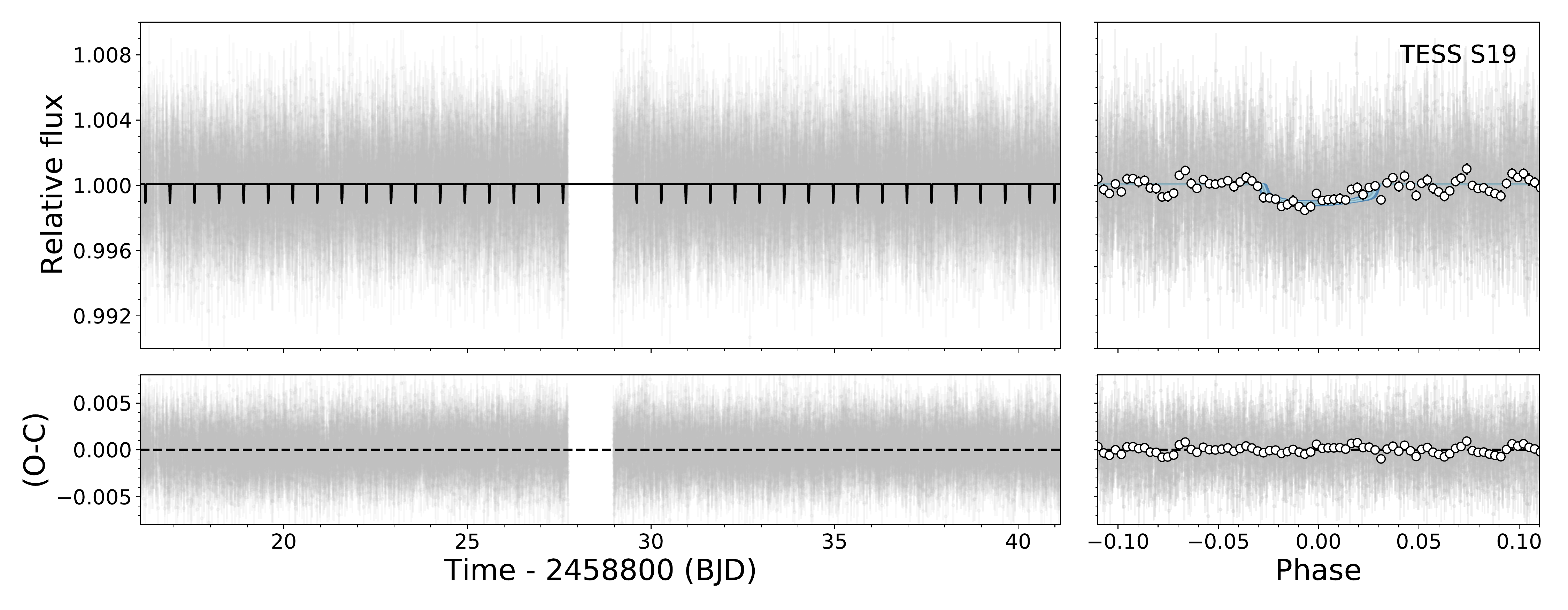}
    \includegraphics[width=\textwidth]{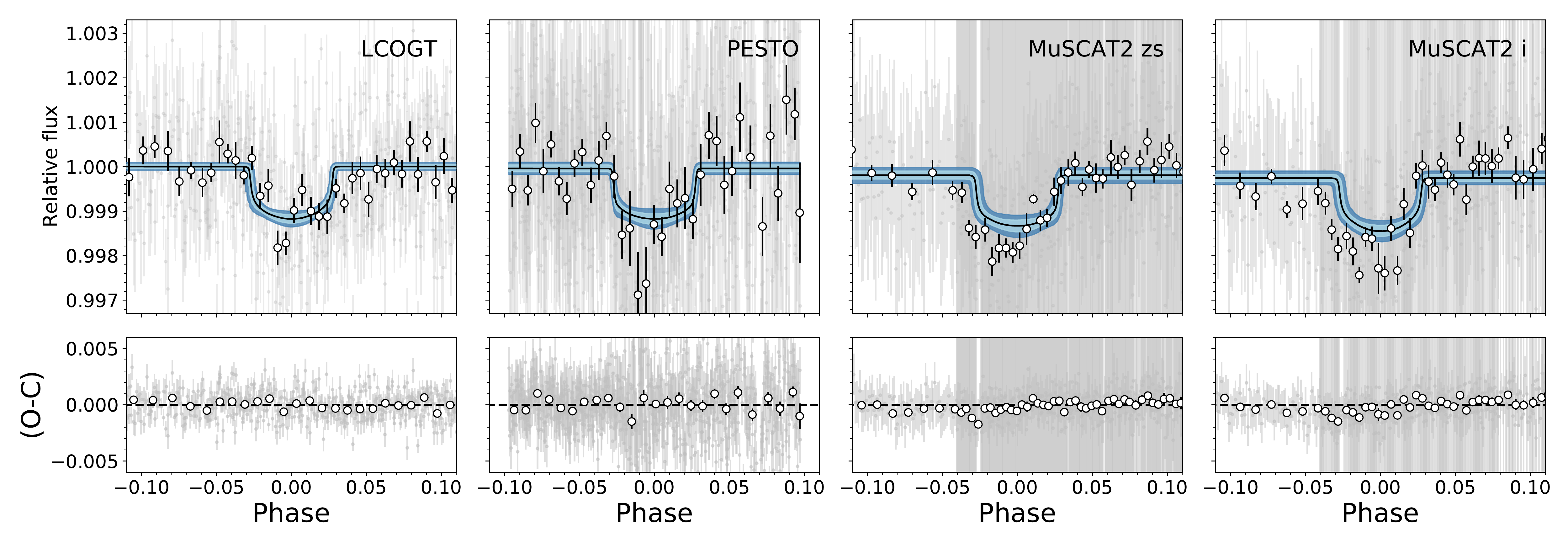}
    \caption{{\em TESS} and ground-based light curves of TOI-1685.
    {\em Upper panels}: PDC simple aperture photometry (SAP) {\em TESS} light curve of TOI-1685 and the corresponding phase-folded light curve.
    {\em Lower panels}: Phase-folded light curves of  LCOGT, PESTO, and MuSCAT2 in filters $i$ and $z_s$ of TOI-1685. For all phase-folded light curves the residuals are shown for the USP found in the system.
    White circles are binned data (shown only for reference; the data used to fit the model were the unbinned points).
    Black curves are the best-fit models, and light and dark blue areas are the 68\,\% and 95\,\% credibility bands, respectively.
    The different sizes of the MuSCAT2 error bars correspond to the two observing nights.}
    \label{fig:joint-fit-photo}
\end{figure*}

Several follow-up studies have suggested that USPs are usually formed in multi-planetary systems \citep{2014ApJ...787...47S}, where multi-body interactions could play an important role in tidal migration. Accurately measuring the masses and orbits of any additional planet in such systems would be helpful in discriminating between different USP origin scenarios.
Thus, in order to understand the processes involved in the formation and evolution of these planets, high cadence photometry and radial velocity (RV) campaigns, able to detect multi-planetary systems, are needed. Because of their short periods, it is relatively easy to precisely measure the parameters of USPs, but it is important to also explore and constrain additional planetary signals in systems that host USPs.

Theory and empirical data have shown that the occurrence rate of small planets tends to increase around late-type stars \citep{2013A&A...549A.109B,2015ApJ...807...45D,2015ApJ...798..112M,2016MNRAS.457.2877G}. 
The {\it Kepler} \citep{,2010Sci...327..977B,KEPLER2016} and {\it K2} \citep{2014PASP..126..398H} space missions uncovered only a few USPs around M dwarfs ($T_{\rm eff} \lesssim$ 4000\,K), such as Kepler-42\,c, Kepler-732\,c, Kepler-32\,b, K2-137\,b, K2-22\,b, and K2-147\,b \citep{2012ApJ...747..144M, 2016ApJ...822...86M, 2018MNRAS.474.5523S, 2017AJ....154..207D, 2018AJ....155..127H}.
However, during the first years of the {\it Transiting Exoplanet Survey Satellite} \citep[{\it TESS};][]{Ricker2015} mission, the number of discoveries  nearly doubled \citep[{LP~791-18\,b, LHS~3844\,b, GJ~1252\,b, LTT~3780\,b, and TOI-1634\,b}; ][]{2019ApJ...883L..16C,2019ApJ...871L..24V, 2020ApJ...890L...7S, 2020A&A...642A.173N, 2021arXiv210312760H}.

In this paper, we report a transiting USP and a potential non-transiting planet candidate around the nearby M3.0 dwarf TOI-1685. 
The USP, with a period of 0.669\,d, was initially discovered as a transiting planet candidate in {\em TESS} data and is confirmed here using ground-based photometry and RV measurements.
The outer non-transiting planet candidate has a longer period of about~9\,d. 

The paper is organized as follows. Section \ref{sec:data} presents the {\it TESS} and ground-based photometry, lucky imaging, and high-resolution spectroscopy of TOI-1685. Section~\ref{sec:star} presents the properties of the host star, either newly derived or collected from the literature. 
In Sect.~\ref{sec:analysis} we present our search for the rotational period of the star, RV modeling, and the joint analysis of all available data made to constrain the properties of the system.
In Sect.~\ref{sec:discussion} we discuss our results and in Sect.~\ref{sec:conclusion} present our conclusions.

\section{Data} \label{sec:data}

\subsection{TESS photometry} \label{sec:tess}

\begin{figure}
    \centering
    \includegraphics[width=0.49\textwidth]{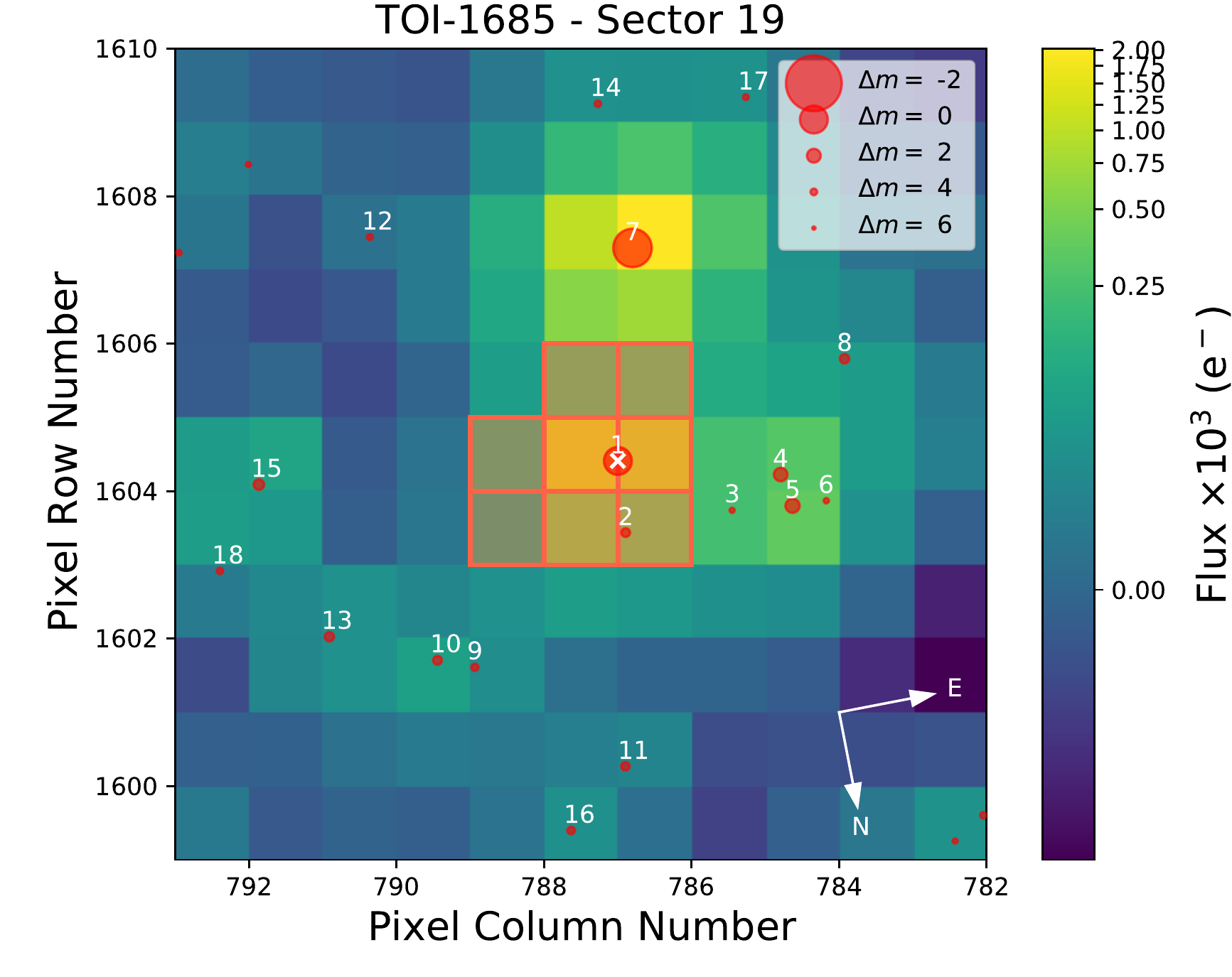}
         \caption{Target pixel files of TOI-1685 in {\em TESS}  sector 19. The electron counts are color coded. The red bordered pixels are used in the SAP. The sizes of the red circles indicate the {\em Gaia} magnitudes of nearby sources down to $G$ = 18\,mag and that of TOI-1685 (circle $\text{}$1 marked with an $\times$).}
        \label{fig:TPF}
\end{figure}

TOI-1685 (TIC~28900646) was observed by {\em TESS} in 2\,min short-cadence integrations during cycle 2 in sector 19 
(see Table~\ref{tab:phot} for details) and was announced on 30~January~2020 as a {\em TESS} object of interest (TOI) through the dedicated {\em TESS} data public website from the Massachusetts Institute of Technology (MIT)\footnote{\url{https://tess.mit.edu/toi-releases/}}.
We downloaded the data from the Mikulski Archive for Space Telescopes\footnote{\url{https://mast.stsci.edu}} (MAST) using the \texttt{lightkurve}\footnote{\url{https://github.com/KeplerGO/Lightkurve}} package \citep{lightkurve}. The photometric light curve was corrected for systematics \citep [Pre-search Data Conditioning (PDC);][]{2012PASP..124.1000S,Stumpe2012PASP..124..985S,Stumpe2014PASP..126..100S}, which is optimized for {\em TESS}  transit searches. 
The upper-left panel of Fig.~\ref{fig:joint-fit-photo} shows the PDC data for {\em TESS} sector 19 with our best-fit model (see Sect.~\ref{subsec:joint} for details).

\begin{table}
\centering
\small
\caption{{\em TESS} observations of TOI-1685.}
\label{tab:phot}
\begin{tabular}{cccll}
\hline\hline
\noalign{\smallskip}
Sector  &  Camera  & CCD   & Start date & End date\\
\noalign{\smallskip}
\hline
\noalign{\smallskip}
19  & 1   &  2  & 28 Nov. 2019 & 23 Dec. 2019\\
\noalign{\smallskip}         
\hline
\end{tabular}
\end{table}

In order to search for any contaminant sources, we placed limits on the dilution factor of TOI-1685. We verified that the sources in the selection aperture in the {\em TESS} target pixel file (TPF) did not significantly affect the depth of the transits. The TPF created with \texttt{tpfplotter}\footnote{\url{https://github.com/jlillo/tpfplotter}} \citep{2020A&A...635A.128A} is shown in Fig.~\ref{fig:TPF}. Within the TPF aperture, we found only one extra source (TIC-28900668, \textit{Gaia} EDR3 252366613254979328), which is separated by 15.6\,arcsec from TOI-1685 and is 3.3\,mag fainter. 
Further information comes from the TOI-1685 {\em Gaia} Early Data Release 3 (EDR3) renormalized unit weight error ({\tt RUWE}) value \citep{2020arXiv201203380L} that is associated with each {\em Gaia} source. 
This is 1.18, below the critical value of 1.40 that indicates that the source may be non-single or otherwise problematic for the astrometric solution.
We estimated the {\em TESS}  minimum dilution factor at $D\sim0.97$ from Eq.~6 in \citet{juliet}. 
Since the PDC light curves are already corrected for possible nearby flux contamination, we fixed this value to 1.0 for all of our model fits presented in the following sections.

\subsection{High-resolution imaging} \label{subsec:lucky}

\begin{figure}
    \centering
    \includegraphics[width=0.49\textwidth]{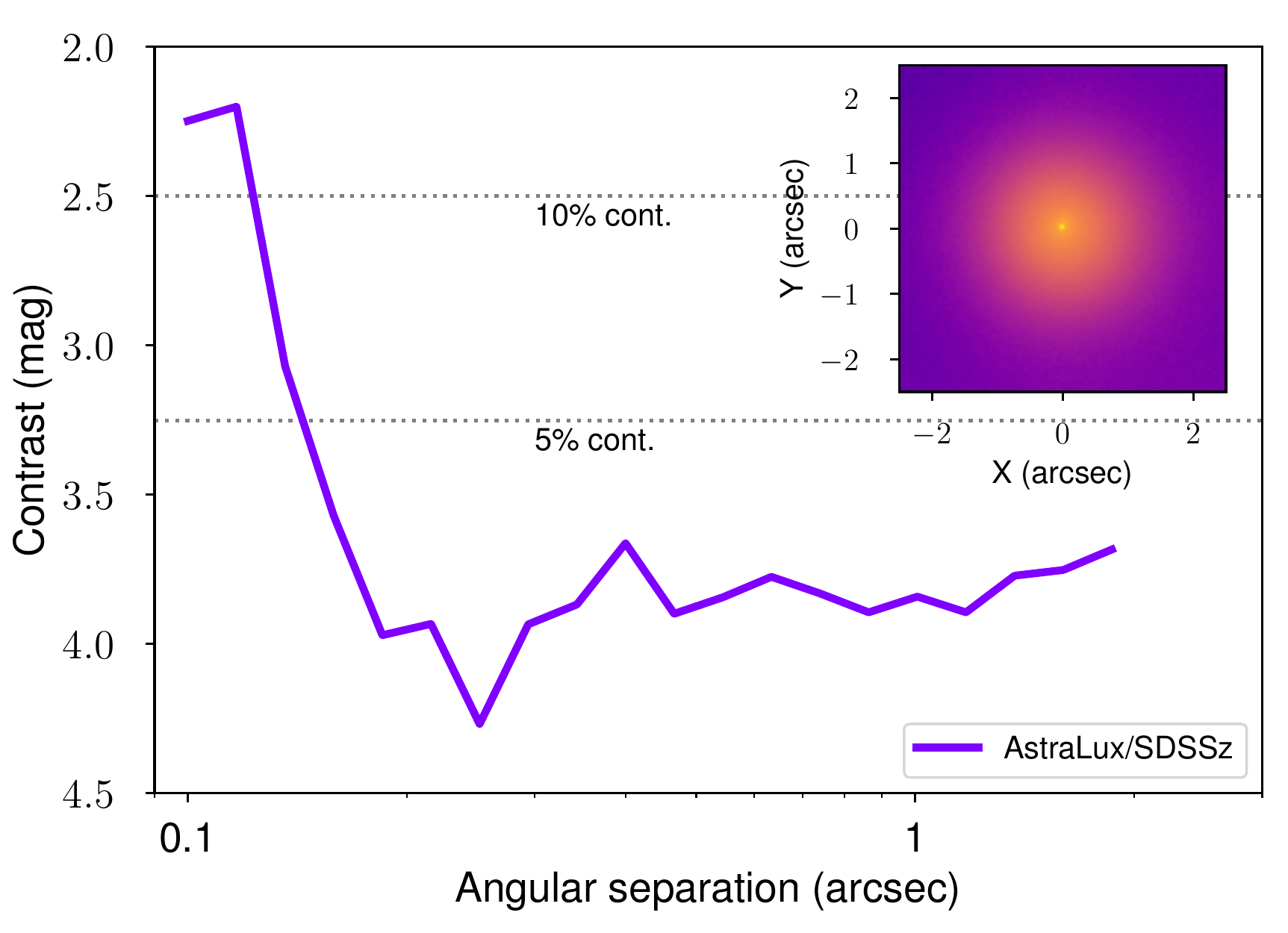}
         \caption{Sensitivity curve for the AstraLux high-spatial-resolution observation of TOI-1685 in the SDSS $z'$ filter. The inset figure shows the AstraLux image when using the best 10\,\% of frames from the whole data cube.}
        \label{fig:Astralux}
\end{figure}

Given the intrinsic faintness of M dwarfs and the large photometric apertures of wide-field surveys ($\sim$21\,arcsec for {\em TESS}), the presence of an unresolved companion must be excluded before a planet candidate is confirmed. 
In some cases, other bright stars in the aperture mask can directly affect the photometry. 
To confirm the identification of the host star and to take nearby potential contaminants into account, we obtained seeing-limited and high-spatial-resolution imaging. 
We also needed to rule out the possibility that the transit in the light curve is due to an eclipsing binary. For this reason, we obtained ground-based photometry.

To search for companions at subarcsecond separations, we observed TOI-1685 with the lucky imaging instrument AstraLux \citep{hormuth08} mounted on the 2.2\,m telescope at the Observatorio de Calar Alto in Almer\'ia, Spain. 
We observed TOI-1685 on 25 February 2020 under good weather conditions with a mean seeing of 1.2\,arcsec and at airmass 1.1. 
The instrument performs imaging with a fast readout (below the coherence time), creating data cubes of thousands of short-exposure frames. Those with the highest Strehl ratio \citep{strehl1902} are subsequently selected and combined into a  final high-spatial-resolution image, which is done by the observatory pipeline \citep{hormuth08}.
We observed in the Sloan Digital Sky Survey (SDSS) $z'$ filter and obtained 87\,600 frames with 20\,ms exposure times and a field of view windowed to $6\times6$\,arcsec. 
Only the best 10\,\%  of the frames were aligned and stacked. 
The final image is shown in the inset panel of Fig.~\ref{fig:Astralux}. Based on this final image, we computed the sensitivity curve using the \texttt{astrasens} package\footnote{\url{https://github.com/jlillo/astrasens}} with the procedure described by \cite{lillo-box12,lillo-box14b}. 
We found no evidence of additional sources within a 2\,$\times$\,2\,arsec$^2$ field of view and within the computed sensitivity limits, as shown in Fig.~\ref{fig:Astralux}. 
This allowed us to set an upper limit to the contamination in the light curve of around 10\,\% down to 0.1\,arcsec.

We further used this contrast curve to estimate the probability of contamination from blended sources in the {\em TESS} aperture that are undetectable in the public images. 
This probability is called blended source confidence (BSC), and the steps for estimating it were described by \cite{lillo-box14b}. 
We used a python implementation of this approach (\texttt{bsc}) 
that uses the {\sc trilegal}\footnote{\url{http://stev.oapd.inaf.it/cgi-bin/trilegal}} Galactic model \citep[v1.6;][]{girardi12} to retrieve a simulated source population of the region around the corresponding target.
This is done in python with the \texttt{astrobase} implementation by \citep{2020zndo...3723832B}. 
This simulation is used to compute the density of stars around a target position (within a radius of $\rho$ = 1\,deg) and derive the probability of a chance alignment at a given contrast magnitude and separation. 
We used the default parameters for the bulge, halo, and disk (thin and thick), as well as the log-normal initial mass function from \cite{chabrier01}.
We applied this technique to TOI-1685. Given the transit depth of planet TOI-1685\,b, this signal could be mimicked by blended eclipsing binaries with magnitude contrasts of up to $\Delta m_{\rm b,max} = 7.5$\,mag\footnote{Maximum contrast (with respect  to  the  measured  flux); see Sect. 4.4.1 and Eq.~2 of \citep{lillo-box14b}.} in the SDSS $z'$ bandpass. 
However, the high-spatial-resolution images provided a low probability of 1.5\,\% for an undetected source with such a magnitude contrast. 
The probability of this source being an appropriate eclipsing binary is well below 0.1\,\%. Given these numbers, we further assumed that the transit signal is not due to a blended binary star and that the probability of a contaminating source is very low.

\subsection{Ground-based seeing-limited photometry}\label{subsec:phot}

\begin{table*}
\caption{Ground-based transit follow-up observations of TOI-1685.}
\label{tab:transitdata}
\centering
\begin{tabular}{lllcccccc}
\hline \hline
\noalign{\smallskip}
Instrument & Country & Date & Filter & Exposure & Duration$^{a}$ & $N_{\rm obs}$ & rms$^{b}$\\
 &  &  & &[s] & [min] &  & [ppt] \\
\noalign{\smallskip}
\hline 
\noalign{\smallskip}
LCOGT   & USA    & 26 August 2020    & ${I}$            & 50 & 173 & 123 & 1.29\\
        &        & 07 November 2020  & ${i_\mathrm{p}}$ & 25 & 266 & 270 & 1.47 \\
        &        & 11 November 2020  & ${i_\mathrm{p}}$ & 50 & 279 & 199 & 1.01 \\
PESTO   & Canada & 08 March 2020     & ${i'}$           & 15 & 187 & 724 & 2.63 \\
MuSCAT2 & Spain  & 19 January 2020   & ${z_\mathrm{s}}$ & 15 & 179 & 328 & 1.24 \\
        &        &                   & ${i'}$           & 15 & 179 & 238 & 1.18 \\
        &        & 29 January 2020   & ${z_\mathrm{s}}$ & 15 & 331 & 333 & 1.70 \\
        &        &                   & ${i'}$           & 15 & 331 & 333 & 1.71 \\  
\noalign{\smallskip}
\hline 
\noalign{\smallskip}
LCOGT$^{c}$ & Spain, USA  & 22--31 December 2020 & ${V}$ & 100 & 39.83 [d] & 20 & 13.7\\
\noalign{\smallskip}
\hline 
\end{tabular}
\tablefoot{
    \tablefoottext{a}{Time span of the observation.}
    \tablefoottext{b}{Root mean square in parts-per-thousand.}
    \tablefoottext{c}{Data used only in Sect.~\ref{subsec:Prot} for determining the stellar rotational period. The duration of the long-term monitoring is in days instead of minutes.}   
    }
\end{table*}

One partial and one full transit of the red dwarf TOI-1685 were observed on 19 and 29 January 2021 with the MuSCAT2 instrument \citep{Narita2019} at the 1.52\,m Telescopio Carlos S\'{a}nchez at Observatorio del Teide, Spain. 
MuSCAT2 is a four-channel imager that performs simultaneous photometry in the $g'$, $r'$, $i'$, and $z_\mathrm{s}$ bands. 
However, the low-quality $g'$ and $r'$ data were discarded from the analysis. 
The exposure times of our observations were 15\,s in each band, and the observations were repeated for at least three times the USP period.
Data reduction and photometric analysis were carried out using the custom-built pipeline for MuSCAT2  \citep{Parviainen2020}. 
The pipeline provides aperture photometry for a set of comparison stars and different aperture sizes. 
From them, the final light curves are chosen after a global optimization that takes into account the transit model and several different sources of systematics from covariates. 
The data obtained on 19 January 2021 were significantly affected by poor weather.

Four additional transit observations of TOI-1685 were taken with the Las Cumbres Observatory Global Telescope (LCOGT), on the night of 26 August 2020 and the nights of 7, 9, and 11 November 2020. 
The night of 9 November was discarded due to the bad quality of the data.
Observations were taken with the 1.0\,m telescopes at McDonald Observatory, USA, which were equipped with 4096\,$\times$\,4096\,pixel SINISTRO cameras, using the $I$ filter on the night in August and the $i_p$ filter on all the nights in November. Exposure times were set to 25, 50, and 50\,s for the nights of 7 November, 11 November, and 26 August, respectively. 
Data reduction and photometric analysis were performed with the dedicated LCOGT {\sc banzai} pipeline and {\sc AstroImageJ}, respectively \citep{Collins2017}. 

Finally, another full transit of TOI-1685 was observed at Observatoire du Mont-M\'egantic, Canada, on 8 March 2020. Using the 1.6\,m telescope equipped with the PESTO camera, the data were obtained in the $i'$ filter with a 15\,s exposure time. The bias subtraction, flat field division, and light curve extraction were also carried out using {\sc AstroImageJ}.

Table~\ref{tab:transitdata} summarizes the three data sets, including the root mean square (rms) of the light curve fluxes.
The resulting transit light curves obtained with MuSCAT2 (in $i$ and $z_\mathrm{s}$), LCOGT, and PESTO are shown in the bottom panels of Fig.~\ref{fig:joint-fit-photo}. 

\subsection{CARMENES RV measurements}\label{subsec:carmenes}

\begin{table}
\centering
\small
\caption{Stellar parameters of TOI-1685.} \label{tab:star}
\begin{tabular}{lcr}
\hline\hline
\noalign{\smallskip}
Parameter   & Value             & Reference \\ 
\hline
\noalign{\smallskip}
\multicolumn{3}{c}{\em Name and identifiers}\\
\noalign{\smallskip}
Name           & 2MASS J04342248+4302148  & 2MASS\\
Karmn$^{a}$    & J04343+430               & AF15 \\
TOI            & 1685                     & ExoFOP-TESS \\  
TIC            & 28900646                 & Sta18 \\  
\noalign{\smallskip}
\multicolumn{3}{c}{\em Coordinates and spectral type}\\
\noalign{\smallskip}
$\alpha$ (J2000)$^{b}$ & 04:34:22.55  & {\it Gaia} EDR3 \\
$\delta$ (J2000)$^{b}$ & +43:02:13.3  & {\it Gaia} EDR3 \\
Sp.\ type         & M3.0\,V      & Terr15 \\
$G$ [mag]        & $12.284\pm2.452$ & {\it Gaia} EDR3 \\
$T$ [mag]        & $11.111\pm0.007$ & Sta19\\
$J$ [mag]        & $9.616\pm0.018$ & 2MASS \\
\noalign{\smallskip}
\multicolumn{3}{c}{\em Parallax and kinematics}\\
\noalign{\smallskip}
$V_r$ [$\mathrm{km\,s^{-1}}]$           & $-43.4\pm8.5$   & Terr15  \\
$\varpi$ [mas]  & $26.589\pm0.019$ & {\it Gaia} EDR3 \\
$d$ [pc]        & $37.609\pm0.027$ & {\it Gaia} EDR3 \\
$\mu_{\alpha}\cos\delta$ [$\mathrm{mas\,a^{-1}}$]  & $+37.762 \pm 0.022$ & {\it Gaia} EDR3 \\
$\mu_{\delta}$ [$\mathrm{mas\,a^{-1}}$] & $-87.062 \pm 0.047$ & {\it Gaia} EDR3 \\
$U$ [$\mathrm{km\,s^{-1}}$]  & $+35.6\pm 8.0$ & This work \\
$V$ [$\mathrm{km\,s^{-1}}$]  & $-29.8\pm2.9$ & This work \\
$W$ [$\mathrm{km\,s^{-1}}$]  & $-3.14\pm0.47$ & This work \\
Gal. population & Thin disk & This work \\
\noalign{\smallskip}
\multicolumn{3}{c}{\em Photospheric parameters}\\
\noalign{\smallskip}
$T_{\mathrm{eff}}$ [K]  & $3434 \pm 51$ & This work   \\
$\log g$                & $4.85\pm 0.04$ & This work   \\
{[Fe/H]}                & $-0.13 \pm 0.16$ & This work   \\
$v \sin i_\star$ [$\mathrm{km\,s^{-1}}$]    & $<2.0$ & This work \\
\noalign{\smallskip}
\multicolumn{3}{c}{\em Physical parameters}\\
\noalign{\smallskip}
$L_\star$ [$L_\odot$]         & $0.0303 \pm 0.0005$ & This work \\
$M_\star$ [$M_{\odot}$]       & $0.495 \pm 0.019$ & This work \\
$R_\star$ [$R_{\odot}$]       & $0.492 \pm 0.015$ & This work \\
\noalign{\smallskip}
\multicolumn{3}{c}{\em Activity and age}\\
\noalign{\smallskip}
pEW(H$\alpha$) [\AA]    & $+0.51\pm 0.06 $ & This work \\
Age (Ga)                &  0.6--2.0     & This work \\
\noalign{\smallskip}
\hline
\end{tabular}
\tablebib{
    2MASS: \citet{2MASS};
    AF15: \citet{2015A&A...577A.128A};
    ExoFOP-TESS: \url{https://exofop.ipac.caltech.edu/tess/};
    {\it Gaia} EDR3: \citet{gaiacollaboration2020gaia};
    Sta18: \citet{2018AJ....156..102S};
    Terr15: \citet{2015ApJS..220...16T};
    Sta19: \citet{2019AJ....158..138S}.
    \tablefoot{ 
    \tablefoottext{a} {Target acronym from the CARMENES input catalog of M dwarfs (see AF15).}
    \tablefoottext{b}{\em Gaia \rm EDR3 equatorial coordinates in equinox J2000 and at epoch J2016.}}
    }
\end{table}

TOI-1685 was observed 55 times with CARMENES between 8~August~2020 and 9~November~2020.
CARMENES \citep{CARMENES, CARMENES18} is a high-resolution spectrograph installed at the 3.5\,m telescope at the Observatorio de Calar Alto, Spain. 
It splits the incoming light into two beams that feed the visual (VIS;~$0.52$--$\SI{0.96}{\micro\metre}$, $\mathcal{R}=94\,600$) and near-infrared (NIR;~$0.96$--$\SI{1.71}{\micro\metre}$, $\mathcal{R}=80\,400$) channels via optical fibers. 
Exposure times ranged between about 1300\,s (limited to the time needed to achieve a signal-to-noise ratio of 150, based on information from real-time exposure meters) and 1800\,s (the maximum exposure time).
We followed the standard data flow of the CARMENES guaranteed time observations \citep{2016SPIE.9910E..0EC}.
In particular, we reduced the spectra with \texttt{caracal} \citep{2014A&A...561A..59Z} and determined the corresponding RVs and spectral activity indices (see Sect.~\ref{subsec:rv_model}) with \texttt{serval} \citep{2018A&A...609A..12Z}.
The RVs were corrected for barycentric motion, instrumental drift, secular acceleration, and nightly zero points (see \citealt{Kaminski18}, \citealt{2019MNRAS.484L...8T}, and, especially, \citealt{2020A&A...636A..74T} for details).

The RVs, activity indices, and their corresponding uncertainties are listed in Table~\ref{tab:RV_Activity_all}. 

\begin{figure*}
    \centering
    \includegraphics[width=\textwidth]{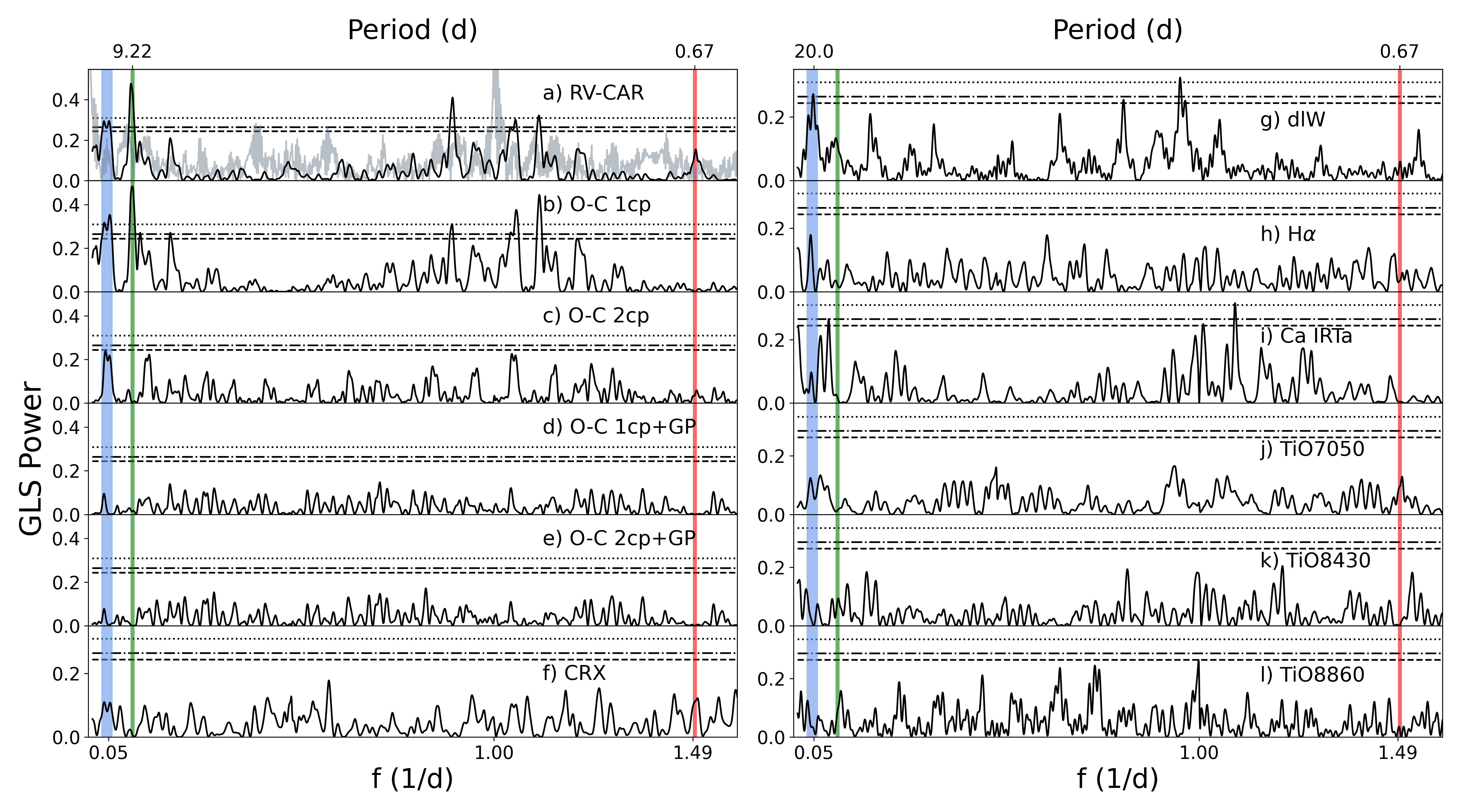}
    \caption{GLS periodograms of:
    ($a$) RVs from CARMENES VIS and the respective spectral window function (gray),
    ($b$) RV residuals after subtracting the one-planet signal,
    ($c$) RV residuals after subtracting the two-planet signal,
    ($d$) RV residuals after subtracting the one-planet signal plus GP,
    ($e$) RV residuals after subtracting the two-planet signal plus GP,   
    ($f$-$l$) CRX, dLW, H$\alpha$, Ca~IRTa (``a'' corresponds to the Ca~{\sc ii} $\lambda$8498.0\,{\AA} line), TiO7050, TiO8430, and TiO8860 activity indices from CARMENES VIS data.
    The ``cp'' in the residual models (panels $a$-$e$) corresponds to circular orbits (for a detailed explanation, see Sect.~\ref{subsec:rv_model}).
In all panels the vertical lines indicate the periods of 0.67\,d (red, USP) and $\sim$9\,d (green, second planet candidate).
    The rotational period of the star (blue line) falls in the region between $\sim$19 and 26\,d.
    The horizontal lines mark the theoretical FAP levels of 1\,\% (dotted), 5\,\% (dash-dotted), and 10\,\% (dashed).
    }    
    \label{fig:GLS_CARMENES}
\end{figure*}

\section{Stellar properties} \label{sec:star}

The star TOI-1685 (2MASS~J04342248+4302148, $V\approx$\,13.3\,mag) is a nearby M3.0\,V star at a distance of approximately 37.6\,pc \citep{2020arXiv201205220B}. It has only been tabulated by a few proper-motion surveys \citep{2011AJ....142..138L,2013MNRAS.435.2161F,2015ApJS..220...16T}.
In this work, we recalculated all stellar parameters for this M dwarf. 
In particular, we measured $T_{\rm eff}$, surface gravity $\log{g}$, and iron abundance [Fe/H] from the stacked CARMENES VIS spectra by fitting them with a grid of PHOENIX-SESAM models, as in \cite{2019A&A...627A.161P},
the rotational velocity $v \sin{i}$ with the cross-correlation method, as in \cite{2018A&A...612A..49R}, and the stellar luminosity, $L_\star$, as in \citet{Cifuentes.et.al.2020}. The stellar radius, $R_\star$, was determined through the Stefan–Boltzmann law, $L_\star = 4 \pi R{_\star}^2 \sigma T_{\rm eff}^4$, and the stellar mass, $M_\star$, using the mass-radius relation derived from main-sequence eclipsing binaries by \citet{2019A&A...625A..68S}.
In particular, we used astro-photometry from {\em Gaia} EDR3 and photometry from
Fourth U.S. Naval Observatory CCD Astrograph Catalog (UCAC4), Two Micron All Sky Survey (2MASS), and All Wide-field Infrared (AllWISE) \citep{2012yCat.1322....0Z,2MASS,Cutri2013}.

We measured the pseudo-equivalent width, pEW(H$\alpha$), a key indicator of stellar activity, on the CARMENES stacked spectrum following \citet{Schf19}. 
In Sect.~\ref{subsec:FAP_rv} we report the search for periodic signals in this and other spectroscopic activity indicators. 
As described in Sect.~\ref{subsec:Prot}, we found that TOI-1685 has a rotation period of around 19\,d. 
Although the star is rotating moderately fast for an early M dwarf, and the Galactic velocities $UVW$ indicate that it belongs to the relatively young thin disk (see Table~\ref{tab:star}), it is not associated with any particular young stellar kinematic group. 
The absence of X-ray emission in {\em ROSAT} observations \cite[First ROSAT X-ray Survey (1RXS) ;][]{1999A&A...349..389V}
and ultraviolet emission in the Galaxy Evolution Explorer ({\em GALEX}) images \citep{2017ApJS..230...24B} is also an indication that the star is not very young. 
Using the gyrochronology relations from \cite{2007ApJ...669.1167B} and \cite{2015MNRAS.450.1787A}, and comparing the rotation period of the star with those of members in open clusters such as Praesepe \citep{2017ApJ...839...92R}, we estimated an age of 0.6--2\,Ga for the system.

Table~\ref{tab:star} summarizes the stellar properties of TOI-1685, providing average values, uncertainties, and the corresponding references.

\section{Analysis and results}\label{sec:analysis}

\subsection{Periodogram analysis of the RV data}\label{subsec:FAP_rv}

We explored the generalized Lomb-Scargle (GLS) periodograms \citep{GLS} of the RVs of TOI-1685. 
The periodogram and window function are shown in panel $a$ of Fig.~\ref{fig:GLS_CARMENES}. 
The strongest signal was found at about 9\,d, with a nominal false alarm probability (FAP) $<$~1\,\%, and its aliases around periods of 1 d (due to the sampling of the data). 
A double peak is visible in the period range of about 19--26\,d with FAP~$<$~5\,\%, while a small isolated peak is discernible at the orbital frequency of TOI-1685\,b. The formal FAP for this feature is $\gtrsim$ 10\,\%.
However, we need to distinguish between an FAP for a peak anywhere in the frequency range of the periodogram and one at a known frequency in the data.
Usually, the FAP is computed by finding the probability that noise creates a peak in the periodogram higher than what is observed over a wide frequency range, typically taken from near zero out to the Nyquist frequency. 
However, in this case there is a signal at the known orbital frequency of the planet, $f_b$.
We need to assess the probability that random data produce more power than what is observed {exactly} at this frequency.

\begin{figure}
    \centering
    \includegraphics[width=0.49\textwidth]{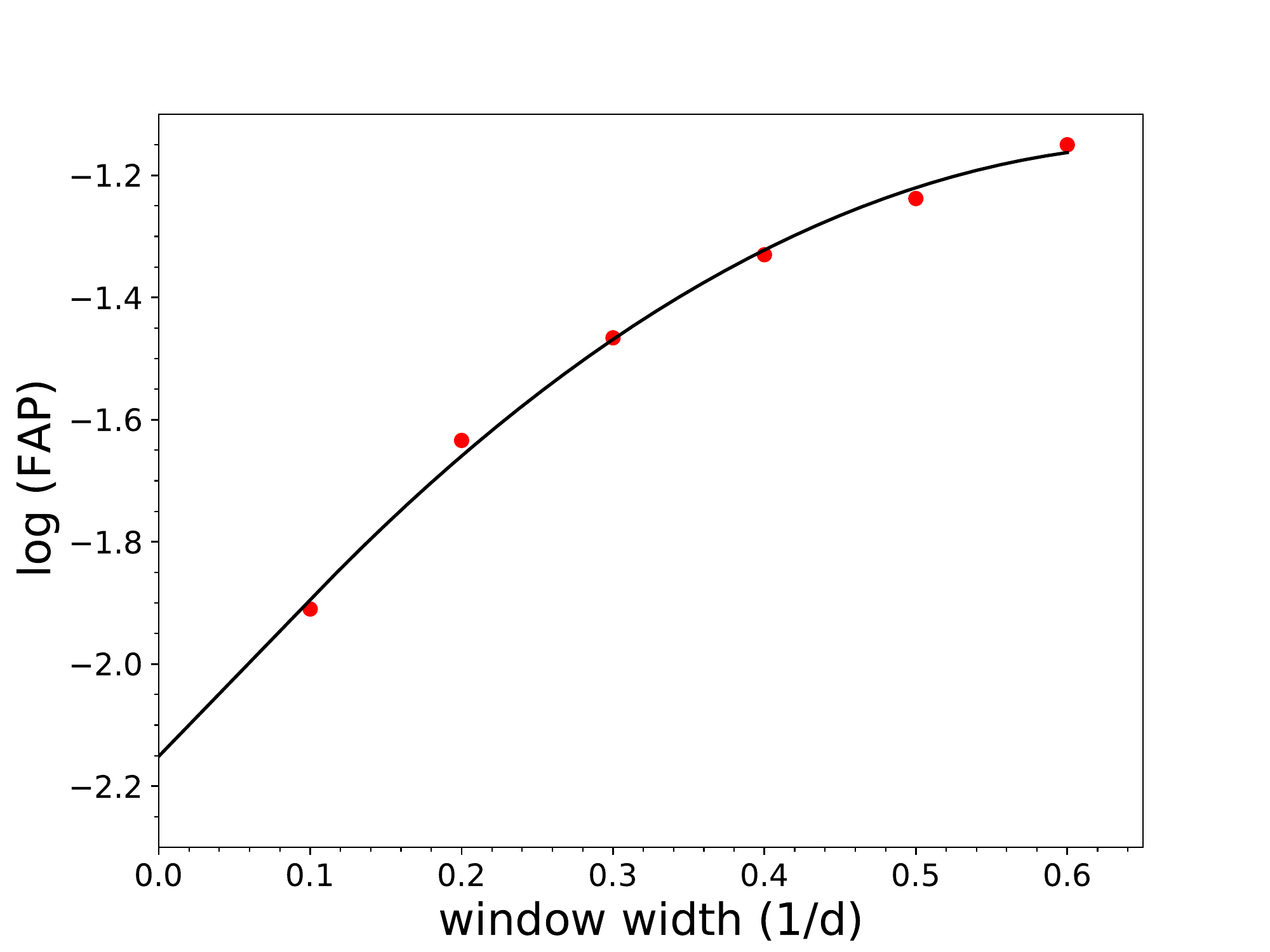}
         \caption{FAP versus window size using a bootstrap as described in Sect.~\ref{subsec:FAP_rv}. Red points are from the bootstrap, and the solid black line is the fit.}
        \label{fig:fap}
\end{figure}

A better estimate of the FAP comes from using the bootstrap randomization method.
Therefore, we randomly shuffled the RV values while keeping the time stamps fixed and noted how often a peak had a power higher than what was observed. 
However, this must be done over a narrow frequency range centered on $f_b$, which can be problematic. 
Too large a window and the FAP is over-estimated, too narrow and the results may not be statistically significant. 
As a result, we employed a "windowing" bootstrap method \citep{2019dmde.book.....H} to compute the FAP over a wide frequency window centered on $f_b$ and then successively narrowed the window for additional bootstraps. 
The fit of the FAP versus window size, extrapolated to zero window length, yields the FAP at $f_b$. This method yielded an FAP $\approx$ 0.007, based on 100\,000 bootstraps, as shown in Fig.~\ref{fig:fap}. This fit confirms that the FAP of the peak at the orbital frequency of the transiting planet is less than 1\,\%.

\subsection{Searching for the rotation period}\label{subsec:Prot}

In order to understand the origin of the $\sim$9\,d  and the double-peak ($\sim$19--26\,d) signals present in the RV data, we searched for additional information in the periodograms of the activity indicators that \texttt{serval} provides, which are shown in panels $f$--$i$ of Fig.~\ref{fig:GLS_CARMENES}. These indicators comprise the chromatic index (CRX), differential line width (dLW), H$\alpha$ line emission, and Ca\,{\sc II} infrared triplet (Ca~IRT) emission. 
The titanium oxide indices that quantify the strengths of the TiO $\gamma$, $\epsilon$, and $\delta$ absorption band heads at 7050\,{\AA}, 8430\,{\AA}, and 8860\,{\AA} were derived from the individual CARMENES spectra following \citep{2018A&A...609A..12Z, Schf19} and are shown in panels $j$, $k$, and $l$, respectively. 
The double-peak signal visible in the RV periodogram is also strong in dLW ($\approx$20\,d; panel $g$), which may indicate that this signal is related to stellar activity \citep{2018A&A...609A..12Z}.

As expected for an early-type M dwarf, the TiO7050, TiO8430, and TiO8860 indices, usually used to measure the properties of cool starspots of magnetically active stars, do not show significant signals. A measured median pEW(H$\alpha$) of +0.51\,\AA\  classifies TOI-1685 as an H$\alpha$ inactive star \citep{2018A&A...614A..76J}. This is consistent with its low $v \sin i_\star$ value of $<$~2.0\,km$\rm s^{-1}$. The activity indices and their uncertainties are listed in Table~\ref{tab:RV_Activity_all}.

\begin{figure}
    \centering
    \includegraphics[width=0.49\textwidth]{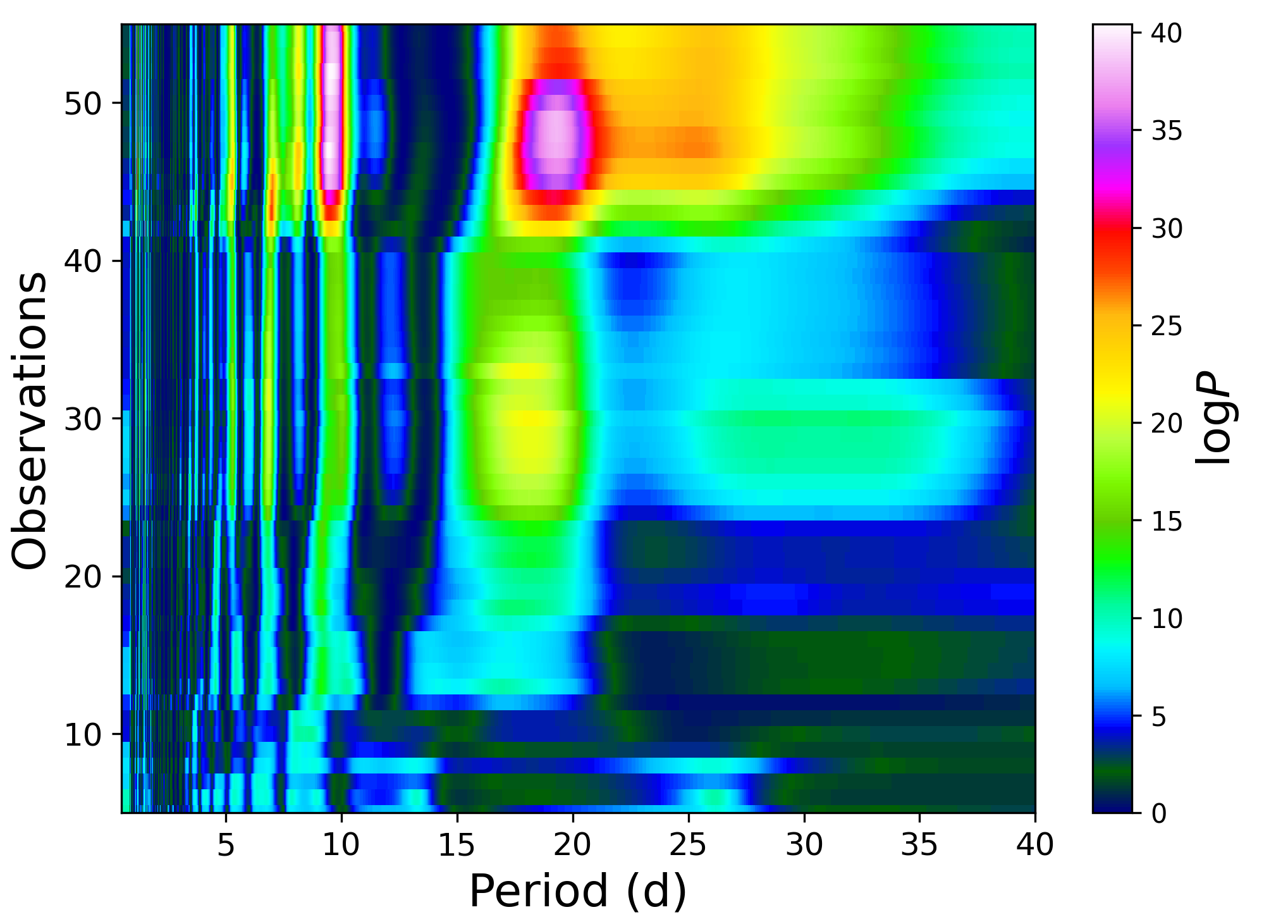}
        \caption{Evolution of the s-BGLS periodogram of the RV data between 0.5\,d and 40\,d, after subtracting the USP signal. The planet candidate is  at $\sim$9\,d, and the activity signal related to the rotational period is visible around 19\,d. The number of data points included in the computation of the periodogram increases from bottom to top.}
        \label{fig: sBGLS_TOI1685}
\end{figure}

In order to explore the stability of the double-peak signal, we computed the stacked-Bayesian GLS (s-BGLS) periodogram of the RV data with the normalization from \cite{sBGLS2017}. The main idea was to stack the RV periodograms by subsequently adding observations and recalculating the periodogram. Figure~\ref{fig: sBGLS_TOI1685} shows an s-BGLS periodogram between 0.5--40\,d, after subtracting the USP signal. The signal at $\sim$9\,d shows a first probability maximum after around 44 observations, after which the probability monotonically increases, as is expected for a Keplerian signal. On the other hand, the s-BGLS of the double-peak signal (centered around $\sim$19\,d) shows a first probability maximum after around 44 measurements and then decreases for some time. This incoherence is characteristic for a non-planetary origin of the signal, and due to the evidence from the dLW we attributed it to the stellar rotation. 

\begin{figure}
    \centering
    \includegraphics[width=0.49\textwidth]{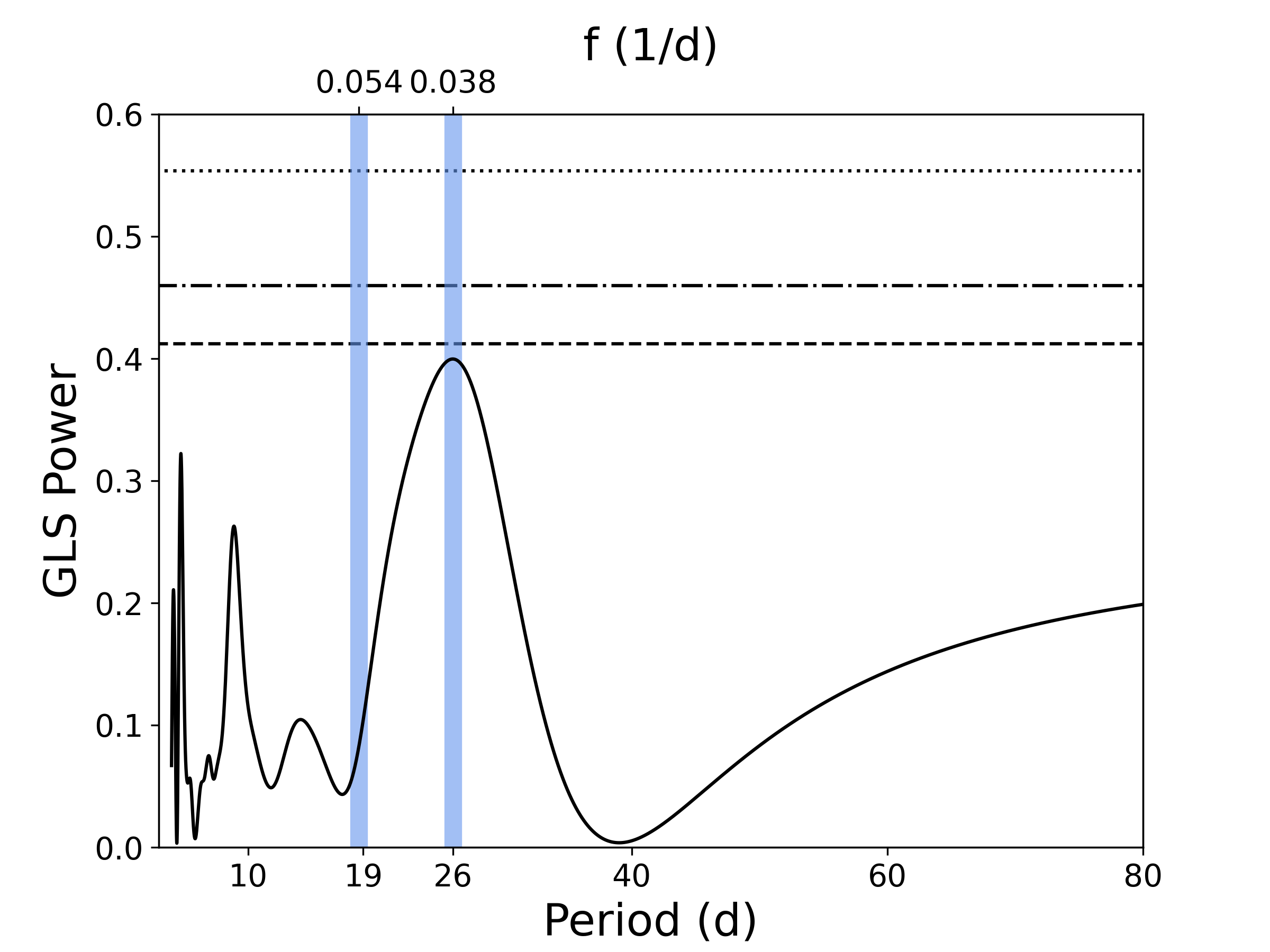}
        \caption{GLS periodogram of the joint $V$-band LCOGT Teide and Haleakal\=a light curve. The horizontal lines mark the theoretical FAP levels of 1\,\% (dotted), 5\,\% (dash-dotted), and 10\,\% (dashed). 
        The highest peak, near the 10\,\%~FAP level, which is related to the rotational period and has a period of $26.0\pm2.5$\,d, and the signal found in the RV data at $\sim$19\,d are marked with blue vertical bands.
        }
        \label{fig: LCO_TOI1685}
\end{figure}

Additionally, we observed TOI-1685 in the $V$ band with the 40\,cm telescopes of LCOGT at the Teide and Haleakal\=a observatories. The 40\,cm telescopes are equipped with 3k\,$\times$\,2k Santa Barbara Instrument Group Charge-coupled device (SBIG CCD) cameras.
with identical pixel scales of 0.571\,arcsec and fields of view of 29.2$\times$19.5\,arcmin.
Weather conditions at both observatories were mostly clear, and the average seeing varied from 2.0\,arcsec to 4.0\,arcsec (for observation details, see Table~\ref{tab:transitdata}).
Raw data were processed with the {\tt banzai} pipeline, which includes bad pixel, bias, dark, and flat-field corrections for each individual night. 
We performed aperture photometry for TOI-1685 and three reference stars of the field and obtained the relative differential photometry between the target and reference stars. 
We adopted an aperture of 16 pixels (9.1\,arcsec), which minimizes the dispersion of the differential light curve. 
Figure~\ref{fig: LCO_TOI1685} shows the GLS periodogram of the joint LCOGT Teide and Haleakal\=a photometric data. 
The highest peak close to the 10\,\%~FAP level has a period of $26.0\pm2.5$\,d,  
which supports the notion that this signal is related to stellar activity and is consistent with the double peak at $\sim$19--26\,d found in the spectroscopic data.

Finally, we searched for available photometric data for TOI-1685. We found some data in the All-Sky Automated Survey for SuperNovae light curves \citep[ASAS-SN;][]{2014ApJ...788...48S,2017PASP..129j4502K}, but unfortunately they were not useful for confirming the rotational period of TOI-1685.

\subsection{Modeling results}
\label{subsec:rv_model}

To model the RV data, we used \texttt{juliet}\footnote{\url{https://juliet.readthedocs.io/en/latest/}}\citep{juliet}, which allows fitting the data at a given prior volume. \texttt{juliet} searches the global posterior maximum based on the evaluation of the Bayesian log-evidence ($\ln{\mathcal{Z}}$), with which one can perform formal model comparisons given the differences in $\Delta\ln{\mathcal{Z}}$.  To select our final model we used the criteria described in \cite{2008ConPh..49...71T}, which consider a difference of $\Delta \ln{\mathcal{Z}}>5.0$ between models as "significant" and of $\Delta \ln{\mathcal{Z}}>2.5$ at "moderate," favoring the former over the latter. Models with $\Delta \ln{\mathcal{Z}}<2$ are indistinguishable, which means none of them are preferred over the others.

\texttt{juliet} calculates the log-evidence via nested sampling algorithms. For the joint fit (see Sect.~\ref{subsec:joint} for details) we used \texttt{dynesty} \citep{dynesty} \footnote{\url{https://github.com/joshspeagle/dynesty}}, and for the RV modeling we used \texttt{MultiNest} \citep{multinest}, which employs the \texttt{PyMultiNest} package \citep{pymultinest}. To model Keplerian RV signals we used \texttt{radvel} \footnote{\url{https://radvel.readthedocs.io/en/latest/}}
\citep{radvel}, and for the Gaussian process (GP) modeling we used \texttt{george}\footnote{\url{https://george.readthedocs.io/en/latest/}} \citep{Ambikasaran2015ITPAM..38..252A}.
For the GP, we selected an exp-sin-squared kernel multiplied by a squared-exponential kernel, also known as the quasi-periodic (QP) kernel, which has the following form:
\begin{equation}
\label{eq:GP}
k(\tau) = \sigma_{\rm GP}^{2}~\exp \left(-\alpha_{\rm GP} \tau^{2}-\Gamma \sin^{2} ({\pi \tau}{P_\textnormal{rot}})\right)\quad
,\end{equation}
where $\sigma_{\rm GP}$ is the amplitude of the GP given in m$\rm s^{-1}$, $\Gamma$ is the amplitude of the GP sine-squared component, $\alpha$ is the square of the inverse length scale of the exponential component of the GP given in $\rm d^{-2}$, $\tau$ is the time lag in days, and $P_{\rm rot}$ is the period of the GP-QP component given in days. 
The GP-QP is a kernel that is widely used to model stellar activity signatures \citep[see, e.g.,][and references therein]{2017chsw.confE..18F,nava2020, 2020A&A...643A.112S, 2020A&A...642A.236K, 2020A&A...639A.132B}. 
The advantage of using a multiplied kernel is due to its exp-sine-squared factor, which enables the modeling of complex periodic signals. 
At the same time, the square-exponential factor allows changes in the periodic function over time, that is, either decreasing or increasing its amplitude. 
This combination is suitable for describing stochastic physical processes occurring in  stars, such as the exponential growth or decay of active regions.

\begin{table}
\centering
\small
\caption{RV model selection.}
\label{tab:models}
\begin{tabular}{l l c c}
\hline\hline
\noalign{\smallskip}
Models & Periods & $\ln{\mathcal{Z}}$  & $\Delta\ln{\mathcal{Z}}$   \\
\noalign{\smallskip}
\hline
\noalign{\smallskip}
1cp         & 0.67                    & --186.609~$\pm$~0.107 &  0.0\\
1cp+GP      & 0.67                    & --177.843~$\pm$~0.014 &  8.77\\
2cp         & 0.67, 9.22              & --177.872~$\pm$~0.073 &  8.74\\
1cp+1kp     & 0.67, 9.31              & --178.501~$\pm$~0.076 &  8.11\\
2cp+GP      & 0.67, 9.03              & --176.609~$\pm$~0.062 &  10.00\\
1cp+1kp+GP  & 0.67, 9.02              & --175.149~$\pm$~0.015 &  11.47\\
3cp         & 0.67, 9.12, 19.83       & --175.464~$\pm$~0.032 &  11.15\\
1cp+1kp+1cp & 0.67, 9.01, 19.94       & --174.369~$\pm$~0.042 &  12.24\\
1cp+1kp+1kp & 0.67, 9.00, 20.27       & --174.791~$\pm$~0.011 &  11.82\\
\noalign{\smallskip}
\hline
\end{tabular}
\tablefoot{Bayesian log-evidence for the different models used for the RV data. In the model names, "cp" corresponds to circular models, "kp" to Keplerian orbits with non-fixed eccentricity, and "GP" to the QP kernel GP described in Sect.~\ref{subsec:rv_model}.}
\end{table}

\begin{figure}
    \centering
    \includegraphics[width=0.49\textwidth]{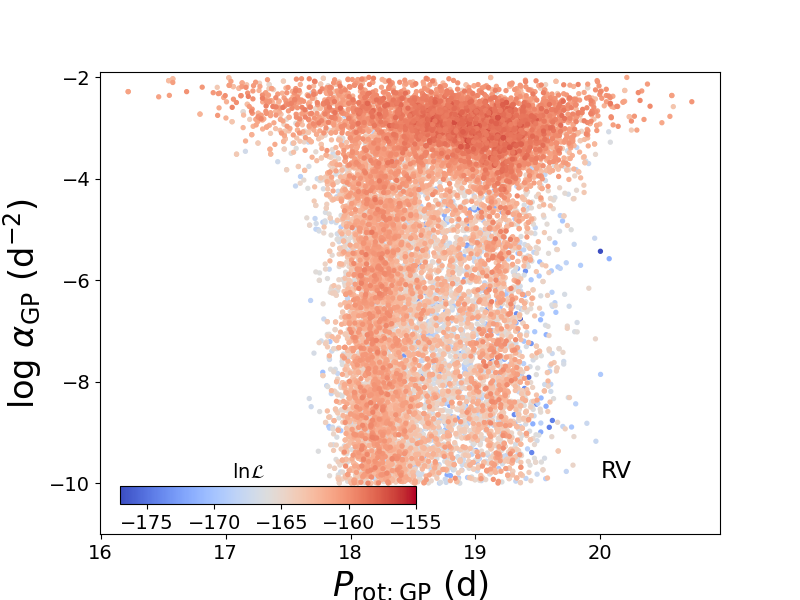}
         \caption{Posterior distribution of the GP fit to the RV data in the $\alpha$ versus $\rm P_{rot}$ plane for TOI-1685. The color coding shows the log-likelihood normalized to the highest value in the posterior sample.}
        \label{fig:alpha_Prot}
\end{figure} 

\subsection{ Only RV data}\label{subsec:rv_inv} 

We performed an extensive model comparison on the RV data to find the model that accounts best for all three signals described in Sect.~\ref{subsec:Prot}. As we discussed, the $\sim$9\,d signal does not seem to be related to the $\sim$19\,d rotational period of TOI-1685; it could be due to a second (non-transiting) planet in the system. 
An overview of the different models and their Bayesian evidence is shown in Table ~\ref{tab:models}.
The residual periodograms for the best log-evidence are shown in Fig.~\ref{fig:GLS_CARMENES}.

Since the USP signal was statistically significant (see Sect.~\ref{subsec:FAP_rv} for details), 
we started fitting the RVs with a one-planet circular model around the USP period using uniform priors between 0.6\,d and 0.7\,d.
The residual periodogram is shown in panel $b$ of Fig.~\ref{fig:GLS_CARMENES}. Here, the strongest periodicity is at 9.22\,d (FAP~$<$~1\,\%). 
After subtracting the USP period and the 9.22\,d signal (using uniform priors between 8\,d and 10\,d) with a circular two-planet fit, only the double-peak signal at $P \approx$~19--26\,d with FAP $\sim$ 10\,\% remained (panel $c$).
The double-peak structure and the activity indicators described in Sect.~\ref{subsec:Prot} show that the $\sim$19--26 \, signal could be related to the stellar rotation period. Therefore, we next investigated whether including a GP to account for this signal improved the log-evidence of the fit.

We did our GP prior selection and final prior volume definition as described by \cite{2020A&A...636A.119S}. 
We started by using a wide prior for the GP~period and GP~$\alpha$ values. 
We constructed a GP $\alpha$-period diagram, which is useful for identifying whether stronger correlated noise (small $\alpha$) favors a certain periodicity \citep[see, e.g.,][and references therein]{2020A&A...643A.112S}. 
With this first approach, the diagram showed a plateau along with all periods in the range $\log{\alpha}\gtrsim$ --2 as well as a structure around the $\sim$19\,d signal. 
The origin of the plateau is that the GP is essentially modeling white noise at that $\alpha$ range.
As we were mostly interested in fitting the spectral region around the suspected stellar rotational period with the GP, we set narrow uniform priors for the signal centered at $\sim$19\,d, and we cut off the plateau by constraining the $\alpha$ values. Figure~\ref{fig:alpha_Prot} shows a scatter plot of the sampled $\alpha$ values of the QP kernel over the sampled rotational periods using the priors presented in Table~\ref{tab:priors}. Considering this plane, we inferred that the likelihood and number of posterior samples around $\sim$19\,d are consistent with a periodic signal present over the entire time of observations.

Once the parameters of our GP-QP were chosen, we performed a simultaneous fit to a one-planet circular model together with a GP (1cp+GP). As we expected, including a GP significantly improved the log-evidence compared to the one-planet circular fit alone ($\Delta\ln{\mathcal{Z}}\sim8.8$). 
To account for the signal at $\sim$9\,d, we further performed a two-planet plus GP model, where we either fixed the eccentricity (2cp+GP) or kept it free (1cp+1kp+GP). In these cases, the differences between these models with the 1cp+GP fit were $\Delta \ln{\mathcal{Z}}$ $\sim$1.2 for 2cp+GP and $\Delta \ln{\mathcal{Z}}$ $\sim$2.7 for 1cp+1kp+GP.
In the first case, the difference made these two models indistinguishable from each other, while in the second the difference made the 1cp+1kp+GP fit moderately favored. 
On the other hand, the difference between them was $\Delta \ln{\mathcal{Z}}$ $\sim$1.5,
which made the models indistinguishable if they were equally likely a priori, so the simplest model should be chosen in this case.

Additionally, we performed a three-planet model fit and compared the $\Delta \ln{\mathcal{Z}}$ with the 1cp+GP fit. 
In this case, we used uniform priors between 15 and 30\,d, the suspected region for the stellar rotational period (Sect.~\ref{subsec:Prot}). In all cases, the differences were $\Delta \ln{\mathcal{Z}}<5$, which implied that none of them were significantly favored. However, we noticed that most of the models that include three signals show a $\Delta \ln{\mathcal{Z}}\gtrsim 2.5$ compared to the 1cp+GP, which makes them moderately favored; hence, we cannot immediately rule out an additional signal in the system.

We decided to choose the 1cp+GP as our fiducial model, and in Sect.~\ref{subsec:Planet_c} we explore the possibility of include the $\sim$9\,d signal with a 2cp+GP model. The residuals of these two fits are shown in panels $d$ and $e$ of Fig.~\ref{fig:GLS_CARMENES}, respectively, where no additional significant peaks are detectable.

\subsection{Joint fit} \label{subsec:joint}

    \begin{table}
    \centering
    \caption{Posterior distributions of the \texttt{juliet} joint fit for the instrumental parameters.}
    \label{tab:posteriors}
    \begin{tabular}{lc} 
        \hline
        \hline
        \noalign{\smallskip}
        Parameter$^{a}$ & TOI-1685 \\
        \noalign{\smallskip}
        \hline
        \noalign{\smallskip}
        \multicolumn{2}{c}{\it Stellar parameters} \\[0.1cm]
        \noalign{\smallskip}
        $\rho_\star$ ($\mathrm{g\,cm\,^{-3}}$)& $5.797^{+0.064}_{-0.060}$\\[0.1 cm]
        \noalign{\smallskip}
        \multicolumn{2}{c}{\it Photometry parameters} \\[0.1cm]
        \noalign{\smallskip}
        $M_{\mathrm{TESS,S19}}$ ($10^{-5}$)   & $-7.63^{+ 1.60}_{- 1.60}$\\[0.1 cm]
        $\sigma_{\mathrm{TESS,S19}}$ (ppm)     & $2.63^{+ 15.44}_{- 2.26}$\\[0.1 cm]
        $q_{1,\mathrm{TESS}}$               & $0.37^{+ 0.27}_{- 0.20}$\\[0.1 cm]
        $q_{2,\mathrm{TESS}}$               & $0.54^{+ 0.29}_{- 0.32}$\\[0.1 cm]
        $M_{\mathrm{LCOGT}}$ ($10^{-5}$)      & $-0.46^{+ 4.31}_{- 4.30}$\\[0.1 cm]
        $\sigma_{\mathrm{LCOGT}}$ (ppm)        & $826.79^{+ 44.54}_{- 42.89}$\\[0.1 cm] 
        $q_{1,\mathrm{LCOGT}}$                 & $0.63^{+ 0.22}_{- 0.29}$\\[0.1 cm]        
        $M_{\mathrm{PESTO}}$ ($10^{-5}$)      & $3.41^{+ 7.36}_{- 7.42}$\\[0.1 cm]
        $\sigma_{\mathrm{PESTO}}$ (ppm)        & $967.07^{+ 22.71}_{- 40.25}$\\[0.1 cm] 
        $q_{1,\mathrm{PESTO}}$                 & $0.49^{+ 0.29}_{- 0.29}$\\[0.1 cm]  
        $M_{\mathrm{MuSCAT2~i}}$ ($10^{-5}$)      & $24.92^{+ 7.97}_{- 7.97}$\\[0.1 cm]
        $\sigma_{\mathrm{MuSCAT2~i}}$ (ppm)        & $7.06^{+ 63.95}_{- 6.49}$\\[0.1 cm] 
        $q_{1,\mathrm{MuSCAT2~i}}$                 & $0.71^{+ 0.20}_{- 0.30}$\\[0.1 cm]    
        $M_{\mathrm{MuSCAT2~z_s}}$ ($10^{-5}$)      & $18.75^{+ 8.84}_{- 8.88}$\\[0.1 cm]
        $\sigma_{\mathrm{MuSCAT2~z_s}}$ (ppm)        & $5.17^{+ 48.83}_{- 4.69}$\\[0.1 cm] 
        $q_{1,\mathrm{MuSCAT2~z_s}}$                 & $0.54^{+ 0.29}_{- 0.32}$\\[0.1 cm]                
        \noalign{\smallskip}
        \multicolumn{2}{c}{\it RV parameters}\\[0.1cm]
        \noalign{\smallskip}
        $\gamma_{\mathrm{CARMENES}}$ ($\mathrm{m\,s^{-1}}$)       & $0.34^{+ 2.69}_{- 2.78}$\\[0.1 cm]
        $\sigma_{\mathrm{CARMENES}}$ ($\mathrm{m\,s^{-1}}$)    & $2.35^{+ 0.79}_{- 0.93}$\\[0.1cm]
        \noalign{\smallskip}
        \multicolumn{2}{c}{\it GP hyperparameters} \\
        \noalign{\smallskip}
        $\sigma_\mathrm{GP,RV}$ ($\mathrm{m\,s^{-1}}$)              & $6.46^{+ 3.17}_{- 1.68}$\\[0.1 cm]
        $\alpha_\mathrm{GP,RV}$ ($10^{-3}\,\mathrm{d^{-2}}$)        & $0.25^{+ 1.70}_{- 0.25}$\\[0.1 cm]
        $\Gamma_\mathrm{GP,RV}$                           & $5.76^{+ 2.75}_{- 3.35}$\\[0.1 cm]
        $P_\mathrm{rot;GP,RV}$ (d)                              & $18.66^{+ 0.71}_{- 0.56}$\\[0.1 cm]
        \noalign{\smallskip}
        \hline
    \end{tabular}
    \tablefoot{
        \tablefoottext{a}{The priors and descriptions for each parameter are given in Table~\ref{tab:priors}. Error bars denote the 68\,\% posterior credibility intervals.}}
\end{table}
\begin{table}
    \centering
    \caption{Derived planetary parameters for TOI-1685\,b and TOI-1685\,[c].}
    \label{tab:derivedparams}
    \begin{tabular}{lc} 
        \hline
        \hline
        \noalign{\smallskip}
        Parameter\tablefootmark{a} & TOI-1685\,b  \\
        \noalign{\smallskip} 
        \hline
        \noalign{\smallskip}
        \multicolumn{2}{c}{\it Planet parameters} \\[0.1cm]
        \noalign{\smallskip}
        $P_{\rm b}$ (d)                              & $ 0.6691403^{+ 0.0000023}_{- 0.0000021}$\\[0.1 cm]
        $t_{0,b}{\rm}$ (BJD)                & $2458816.22615^{+ 0.00059}_{- 0.00060}$\\[0.1 cm]
        $a_{\rm b}/R_\star$                                  & $5.158^{+ 0.019}_{- 0.018}$ \\[0.1 cm]
        $p = R_{\rm b}/R_\star$                                  & $0.0317^{+ 0.0009}_{- 0.0010}$ \\[0.1 cm]
        $b = (a_{\rm b}/R_\star)\cos i_{\rm b}$                                  & $0.473^{+ 0.048}_{- 0.054}$ \\[0.1 cm]
        $i_{\rm b}$ (deg)   & $84.74^{+ 0.60}_{- 0.54}$ \\[0.1 cm]
        $r_1$                                & $0.65^{+ 0.03}_{- 0.04}$ \\[0.1 cm]
        $r_2$                                & $0.0317^{+ 0.0009}_{- 0.0010}$\\[0.1 cm]
        $K_{\rm b}$ ($\mathrm{m\,s^{-1}}$)           & $4.41^{+ 0.73}_{- 0.73}$\\[0.1 cm]
        \noalign{\smallskip}
        \multicolumn{2}{c}{\it Derived physical parameters} \\[0.1cm]
        \noalign{\smallskip}
        $M_{\rm b}$ ($M_\oplus$)    & $3.78 ^{+ 0.63}_{- 0.63}$    \\[0.1 cm]
        $R_{\rm b}$ ($R_\oplus$)        & $1.70^{+ 0.07}_{- 0.07}$   \\[0.1 cm]
        $\rho_{\rm b}$ (g cm$^{-3}$)             & $4.21^{+ 0.95}_{-0.82}$      \\[0.1 cm]
        $g_{\rm b}$ (m s$^{-2}$)                 & $12.79 ^{+ 2.49}_{- 2.28}$     \\[0.1 cm]
        $T_\textnormal{eq,b}$ (K)\tablefootmark{b}          & $1069^{+16}_{-16}$    \\[0.1 cm]
        $S_{\rm b}$ ($S_\oplus$)            & $217^{+ 13}_{-13}$  \\[0.1 cm]
        \noalign{\smallskip}
        \hline
        \noalign{\smallskip}
        Parameter\tablefootmark{b} & TOI-1685\,[c]$^{c}$  \\
        \noalign{\smallskip}
        \hline
        \noalign{\smallskip}
        $P_{\rm c}$ (d)                              & $ 9.025^{+ 0.104}_{- 0.119}$\\[0.1 cm]
        $t_{0,c}{\rm}$ (BJD)                & $2458820.4^{+2.9}_{-2.6}$\\[0.1 cm]
        $a_{\rm c}/R_\star$                                  & $29.23^{+ 0.23}_{- 0.24}$ \\[0.1 cm]
        $K_{\rm c}$ ($\mathrm{m\,s^{-1}}$)           & $4.53^{+ 1.01}_{- 1.02}$\\[0.1 cm]
        $T_\textnormal{eq,c}$ (K)\tablefootmark{c}          & $ 449.1^{+ 6.9}_{-6.9}$    \\[0.1 cm]
        $S_{\rm c}$ ($S_\oplus$)            & $6.76 ^{+ 0.42}_{-0.41}$  \\[0.1 cm]
        $M_{\rm c}\rm sin i$ ($M_\oplus$)        & $9.21^{+ 2.06}_{- 2.07}$    \\[0.1 cm]
        \noalign{\smallskip}
        \hline
    \end{tabular}
    \tablefoot{
      \tablefoottext{a}{Parameters obtained with the posterior values from Table~\ref{tab:posteriors}.
      \tablefoottext{b}{Parameters obtained with the posterior values of the 2cp+GP model fit.}
      \tablefoottext{c}{We use square brackets to denote the tentative second planet as there is no agreed-upon nomenclature for unconfirmed planet candidates. It is understood that the values tabulated here have been computed under the assumption that the RV signal is indeed of a planetary origin.}      
      Error bars denote the 68\,\% posterior credibility intervals.}
      \tablefoottext{c}{The equilibrium temperature was calculated assuming zero Bond albedo.}
      }
\end{table}

\begin{figure*}
    \centering
    \includegraphics[width=\textwidth]{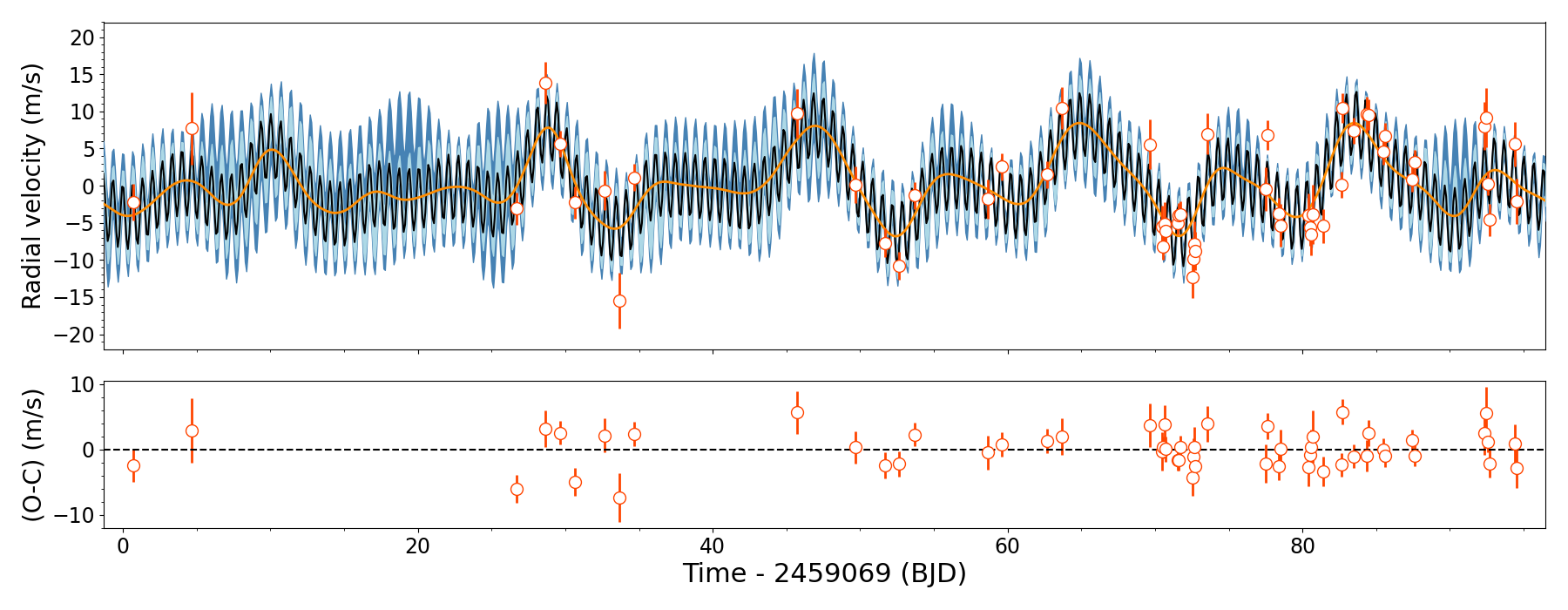}
    \caption{RV data from CARMENES (red). The gray curve is the median best-fit  Keplerian \texttt{juliet} model, the light and dark blue areas are its 68\,\% and 95\,\% credibility bands, and the orange curve is the QP kernel (GP-QP).}
    \label{fig:joint-fit-rv}
\end{figure*}

\begin{figure}
    \centering
    \includegraphics[width=0.49\textwidth]{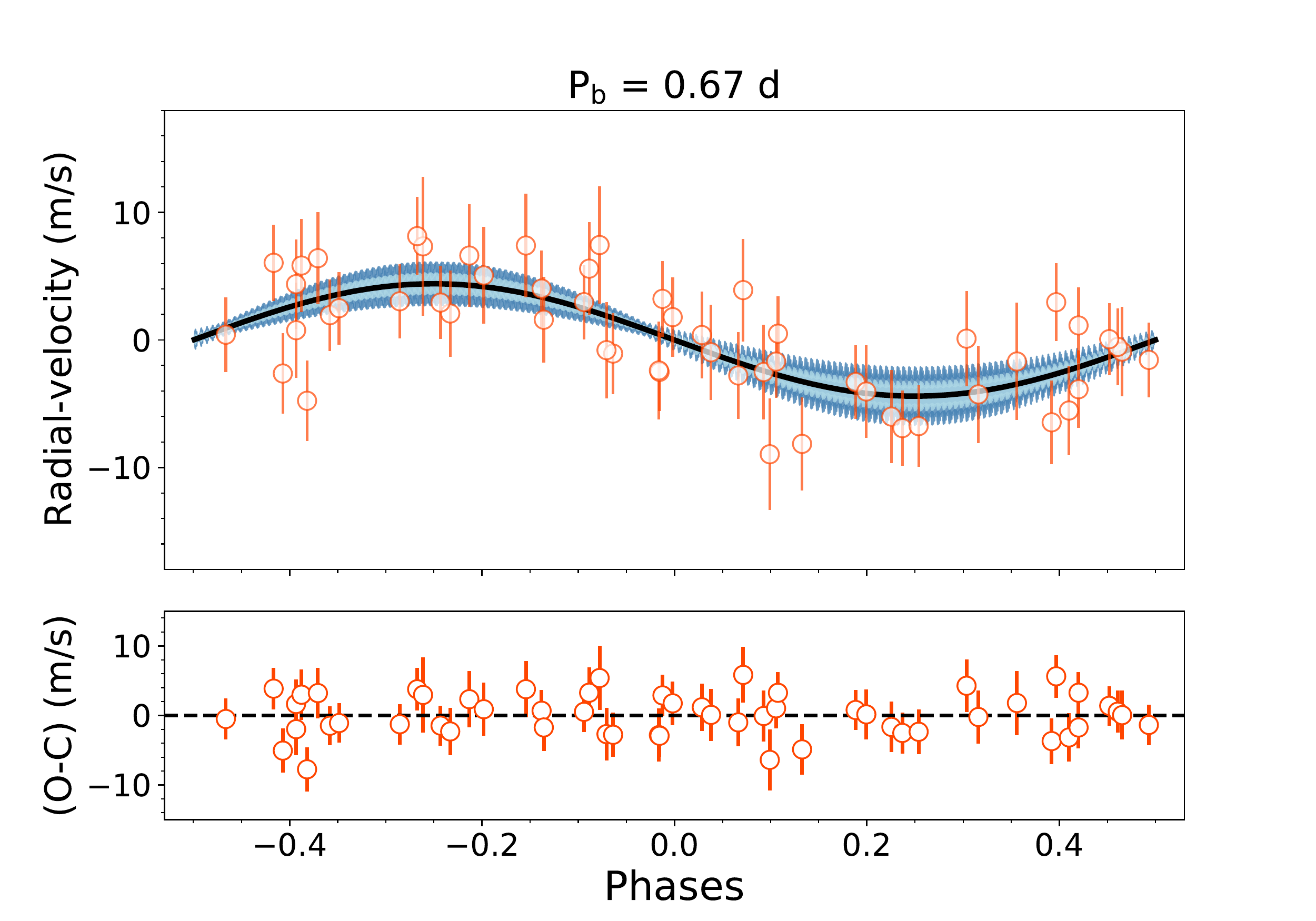}
         \caption{Phase-folded RV curves and their residuals of the USP. Black curves are the best-fit models, and blue areas are the 68\,\% and 95\,\% credibility bands.}
        \label{fig:rv_phase}
\end{figure} 

In order to obtain precise parameters of the TOI-1685 system, we performed a joint analysis with \texttt{juliet}. For the joint fit we used {\it TESS}, LCOGT, PESTO, MuSCAT2, and CARMENES VIS data.
For the transit modeling, \texttt{juliet} makes use of the \texttt{batman} package \citep{batman}. 
To parameterize the quadratic limb-darkening effect in the {\em TESS}  photometry, we employed the efficient, uninformative sampling scheme of \cite{kipping2013} and a quadratic law.
For LCOGT, PESTO, and MuSCAT2 photometry,  we used a linear law to parameterize the limb-darkening effect.
We followed the \cite{Espinoza18} parameterization to explore the full physically plausible parameter space for the planet-to-star radius ratio, $p = R_{\rm p}/R_\star$, and the impact parameter, $b$. 
The model selection was performed based on the analyses on the photometric data plus the highest peaks in the RV periodogram. 
As discussed in Sect.~\ref{subsec:rv_model}, we selected as our fiducial model one planet with a circular orbit and a QP GP for the stellar rotation (1cp+GP). 

The selected priors for our joint fit are presented in Table~\ref{tab:priors}.
The posterior distributions of our joint fit are presented in Tables~\ref{tab:posteriors} and~\ref{tab:derivedparams}, while the resulting photometry and RV models are presented in Figs.~\ref{fig:joint-fit-photo}, ~\ref{fig:joint-fit-rv}, and ~\ref{fig:rv_phase}, respectively. 
The obtained posterior probabilities are presented in Fig.~\ref{fig:corner_plot}.
The maximum posterior of the rotational period of the GP periodic component was around 19\,d, in agreement with the region at $\approx$~19--26\,d observed in the GLS RV periodogram (Fig.~\ref{fig:GLS_CARMENES} and  Sect.~\ref{subsec:Prot}).

\section{Discussion} \label{sec:discussion}

\subsection{Ultra-short-period planet: TOI-1685~b}\label{subsec:USP}

We present the discovery of the USP TOI-1685\,b, which orbits its host star with a period of 0.669\,d. 
To confirm the planetary nature of the TESS transiting candidate, we obtained high-resolution spectra using the CARMENES spectrograph.
We derived a mass of $M_{\rm b}$~=~$3.78 ^{+ 0.63}_{- 0.63}$\,$M_\oplus$, a radius of $R_{\rm b}$~=~$1.70^{+ 0.07}_{- 0.07}$\,$R_\oplus$, and a bulk density of $\rho_{\rm b}$~=~$4.21^{+0.95}_{-0.82}$\,g\,cm$^{-3}$ (see~Table~\ref{tab:derivedparams}).

\begin{figure}
    \centering
    \includegraphics[width=0.49\textwidth]{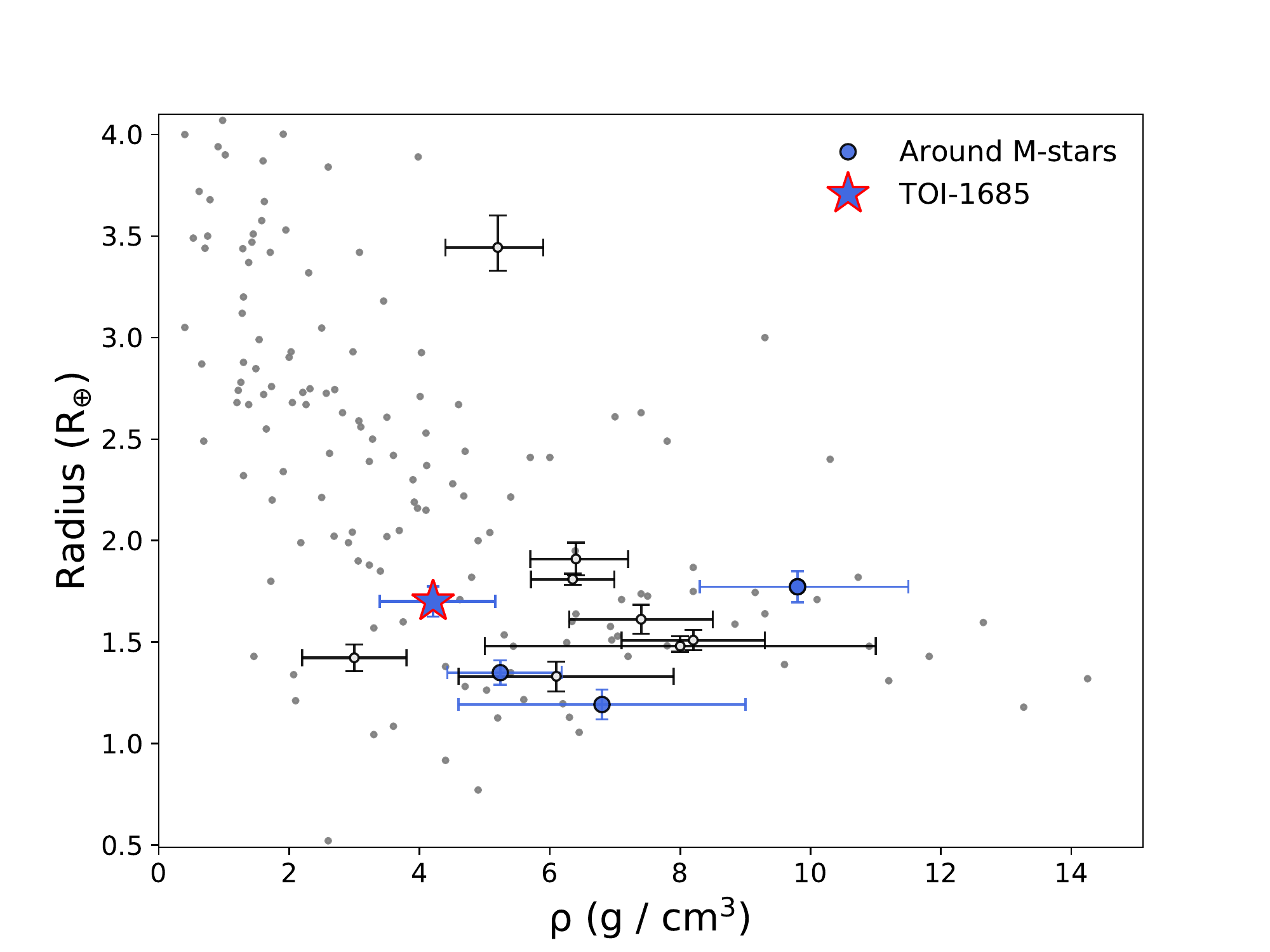}
         \caption{Density-radius diagram for all known exoplanets with planet bulk densities $\rho \le$ 15\,g\,cm$^{-3}$ and radii $R \leq$ 4\,$R_{\oplus}$ (gray) from the NASA exoplanet archive. USPs are drawn in black and are shown with the error bars of the measurements. Blue-filled circles mark USPs around M dwarfs, and the filled star symbol is TOI-1685\,b.}
        \label{fig:planet_comparison}
\end{figure}

Figure~\ref{fig:planet_comparison} shows TOI-1685\,b in the context of all known exoplanets from NASA's exoplanet archive\footnote{\url{https://exoplanetarchive.ipac.caltech.edu/}, \\ \url{http://exoplanet.eu/}}, with $R\leq$ 4\,$R_{\oplus}$ and a planet bulk density of $\rho \le$ 15\,g\,cm$^{-3}$. Here, the USPs with orbital periods ranging from less than 10 hours to about one day tend to be smaller than 2\,$R_{\oplus}$ \citep{2018NewAR..83...37W} and are believed to have lost their atmospheres due to X-ray and ultraviolet (XUV) photo-evaporation from their host stars \citep[e.g.,][]{owen2013,Lopez2013,Jin2014,Chen2016,Owen2017}. With an equilibrium temperature of $T_\textnormal{eq}$~=~$1069\pm16$\,K, it is likely that TOI-1685\,b has gone through a similar process. 
In terms of separation from its host star, TOI-1685\,b is one of the closest known planets with a mass determination. 
An insolation flux of $S$~=~$217\pm13$\,$S_\oplus$ also makes TOI-1685\,b one of the hottest transiting super-Earth discovered to date. TOI-1685\,b is the third known USP to be found orbiting an M star, and the least dense of the three. 

\begin{figure}
    \centering
    \includegraphics[width=0.49\textwidth]{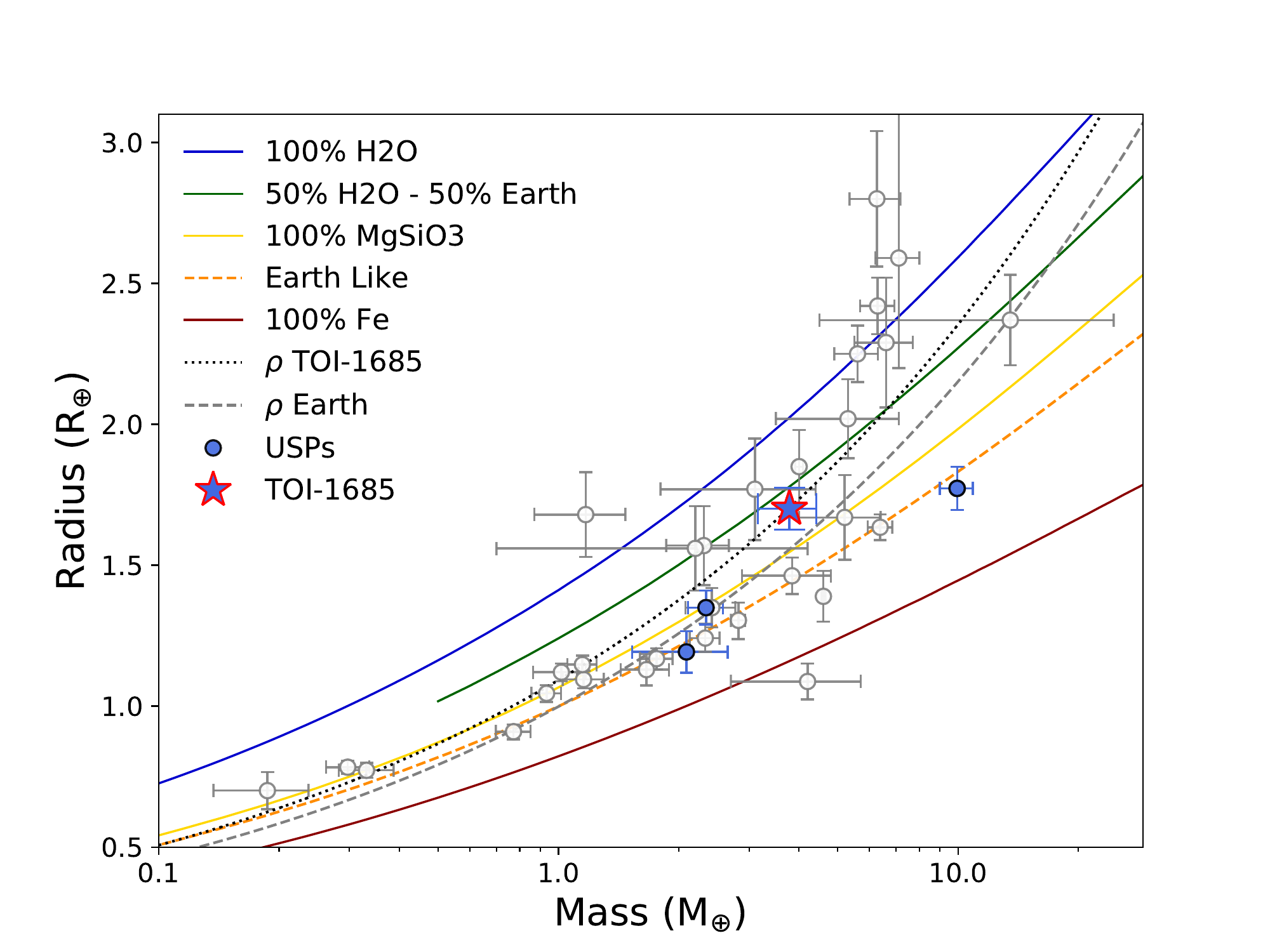}
         \caption{Mass-radius diagram in Earth units. Open circles are transiting planets around M-dwarf stars from the Transiting Extrasolar Planets Catalogue (TEPCat) database (\citealt{2011MNRAS.417.2166S} and \citealt{2019ApJ...887..261M}), blue-filled circles are USPs, and the filled star symbol is TOI-1685\,b.}
        \label{fig:composition}
\end{figure}

A comparison of the physical properties of TOI-1685\,b with compositional models from \citet{2016ApJ...819..127Z, Zeng2019} is shown in Fig. \ref{fig:composition}. The diagram reveals that TOI-1685\,b is consistent with a bulk composition of 50\,\% H$_2$0 and 50\,\% silicate. 

Finally, the proximity of TOI-1685\,b to its host star, and assuming that the planet has not lost its atmosphere, makes TOI-1685\,b an attractive target for atmospheric characterization. In order to estimate the suitability of TOI-1685\,b for such characterization, we calculated the spectroscopic metrics from \citet{Kempton2018}. 
The transmission spectroscopy metric (TSM) and the emission spectroscopy metric (ESM) are analytic metrics for the expected S/N of transmission and emission spectroscopy observations by James Webb Space Telescope ({\em JWST}). 

\begin{figure}
   \centering
   \includegraphics[width=\hsize]{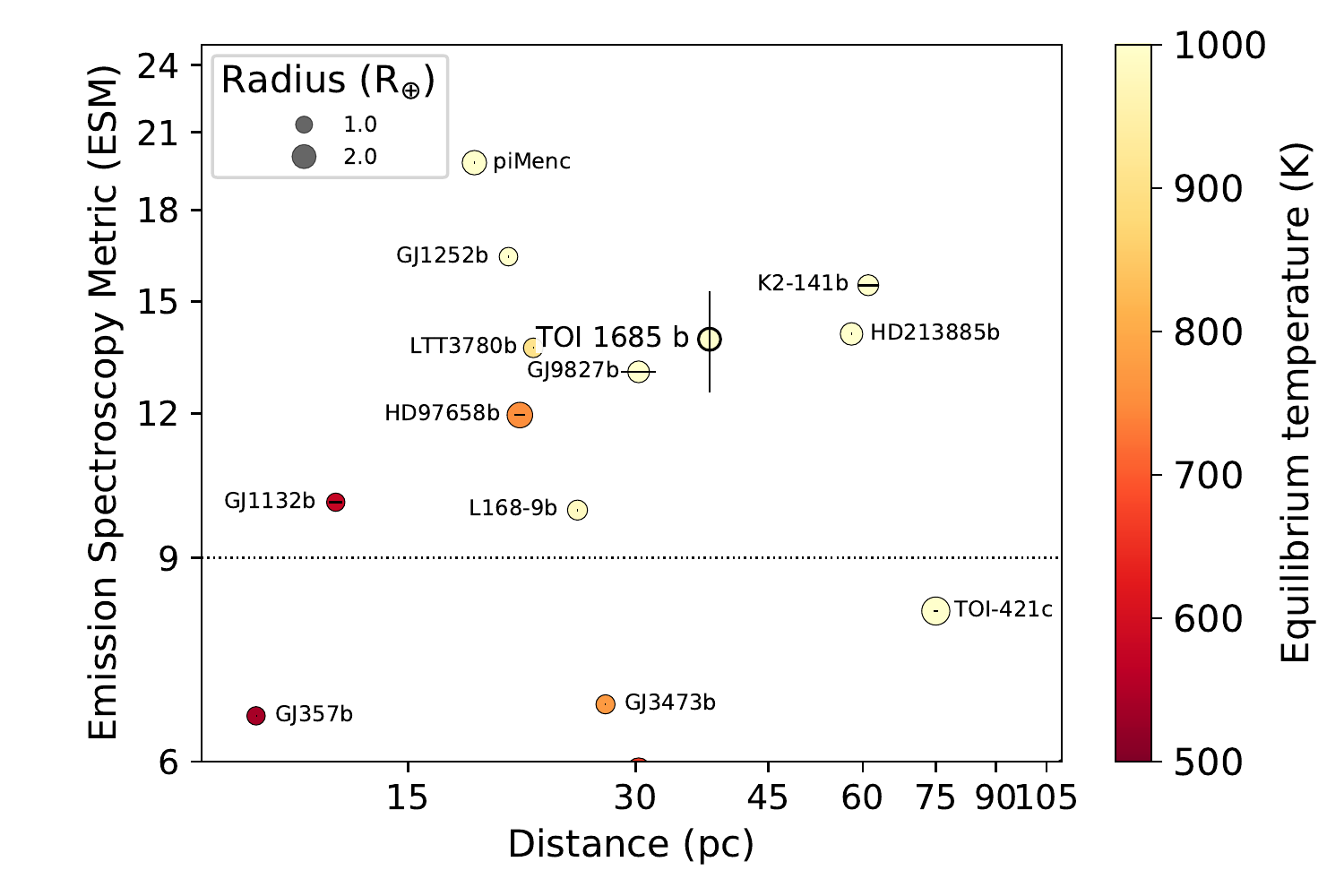}
   \includegraphics[width=\hsize]{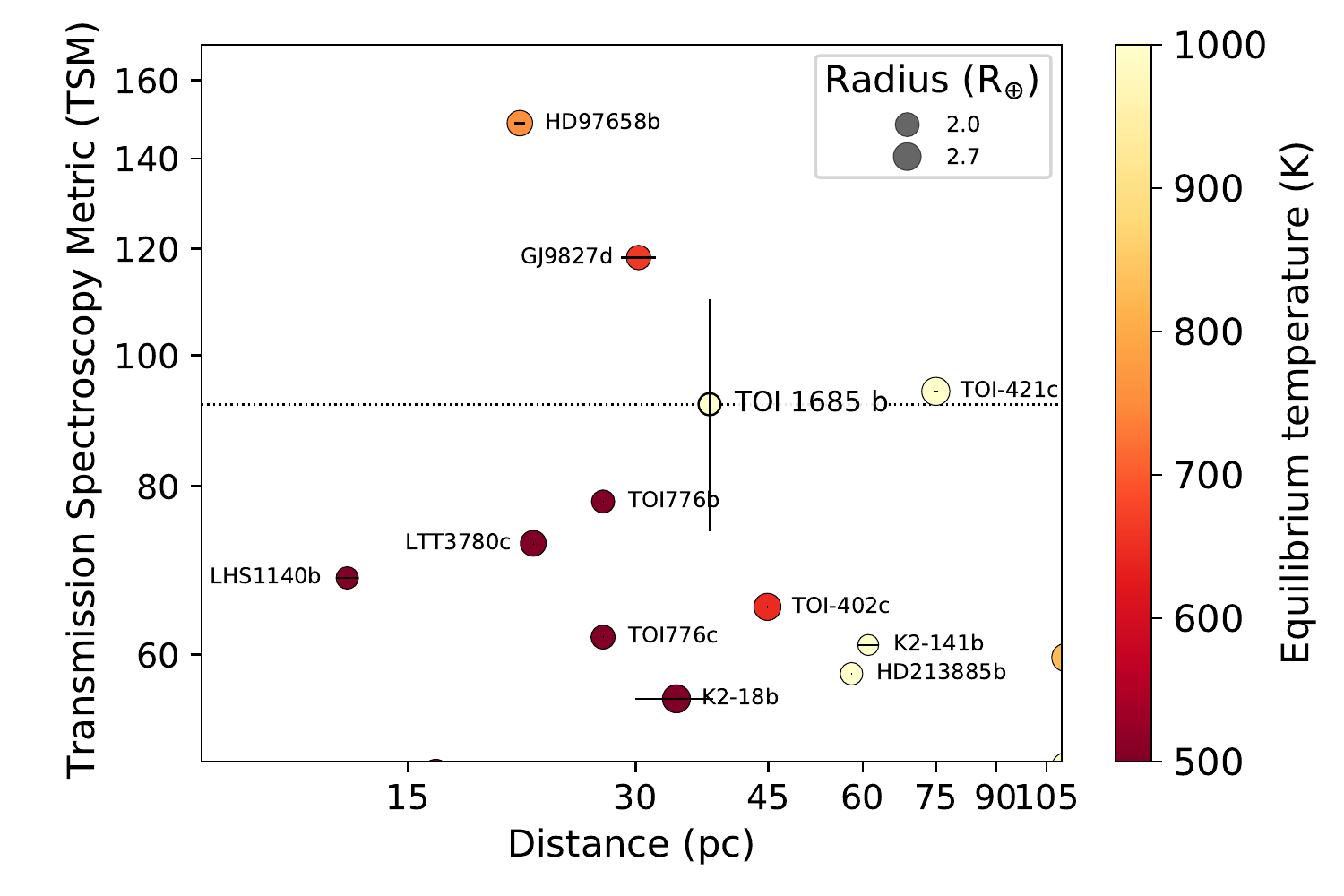}
      \caption{ 
      ESM and TSM metrics for TOI-1685\,b. {\em Top}: ESM for exoplanets with a radius of less than 3\,$R_\oplus$.
      {\em Bottom}: TSM for exoplanets with a radius of less than 3~$R_\oplus$ but larger than 1.5~$R_\oplus$. 
      All planets have mass determination by either RVs or TTVs. 
      TOI-1685\,b is labeled and marked with a thicker black borderline in both panels. For most exoplanets, the error bars, in distance, are smaller than the symbol size. The uncertainties in the ESM and TSM  are only plotted for TOI-1685\,b.
      }
         \label{fig:TSM_ESM}
\end{figure}

The TSM is estimated based on the strength of spectral features and the brightness of the host star, assuming a cloud-free atmosphere. The ESM is an approximation of the expected S/N for a single secondary eclipse observation integrated over the full 5--10\,$\mu$m bandpass of the low-resolution spectroscopy (LRS) mode of the {\em JWST} Mid-Infrared Instrument (MIRI). We estimated the ESM of TOI-1685\,b to be about 13.9. This is larger than that of Gl~1132\,b, which is considered a benchmark rocky planet for emission spectroscopy \citep{Kempton2018}. The top panel of Fig.~\ref{fig:TSM_ESM} shows the ESMs of exoplanets with measured masses, either through RVs or transit-timing variations (TTVs), with a radius from the NASA exoplanet archive\footnote{\url{https://exoplanetarchive.ipac.caltech.edu/}, \\ \url{http://exoplanet.eu/}} of less than 3\,$R_\oplus$. 
We chose this radius cutoff in order keep only the most likely terrestrial planets,  and we excluded potential small sub-Neptunes \citep{Kempton2018}. 
Planets with ESMs on the order of or above the value of Gl~1132\,b are separated from the others by a horizontal dotted line. 
TOI-1685\,b is one of the hottest members of this family of small rocky planets suitable for emission spectroscopy.

We calculated a TSM value of 86$\pm$18 for TOI-1685\,b. The TSMs of small exoplanets (1.5\,$R_\oplus < R_p < 3\,R_\oplus$) are shown in the bottom panel of Fig.~\ref{fig:TSM_ESM}. We excluded planets with radii smaller than 1.5\,$R_\oplus$ from this panel to make the TSMs comparable, as defined in \citep{Kempton2018}. A favorable TSM value for this class of planets is around 90 or higher \citep[Table~1 in][]{Kempton2018}. This implies that TOI-1685\,b would be a suitable target for atmospheric characterization through transmission spectroscopy as well. The suitability of TOI-1685\,b for both transmission and emission spectroscopy makes this planet a worthy target for atmospheric characterization over a wide orbital phase.

The equilibrium temperature of TOI-1685\,b is estimated to be about 1070\,K. 
This is larger than the 880\,K temperature threshold above which planets are expected to have molten surfaces, such as 55~Cnc\,e \citep{mcarthur2004detection}. No thick H$_{2}$-dominated primary atmosphere is expected at these high temperatures, except possibly an exosphere maintained by vaporized rocks \citep{mansfield2019identifying} or a secondary outgassing atmosphere due to volcanic activity. 
If such a substantial exosphere exists, it could provide critical observable tracers to shed light on the planet formation and evolution of USPs as it would directly trace the surface or near-surface composition of these planets.

\begin{figure}
   \centering
   \includegraphics[width=\hsize]{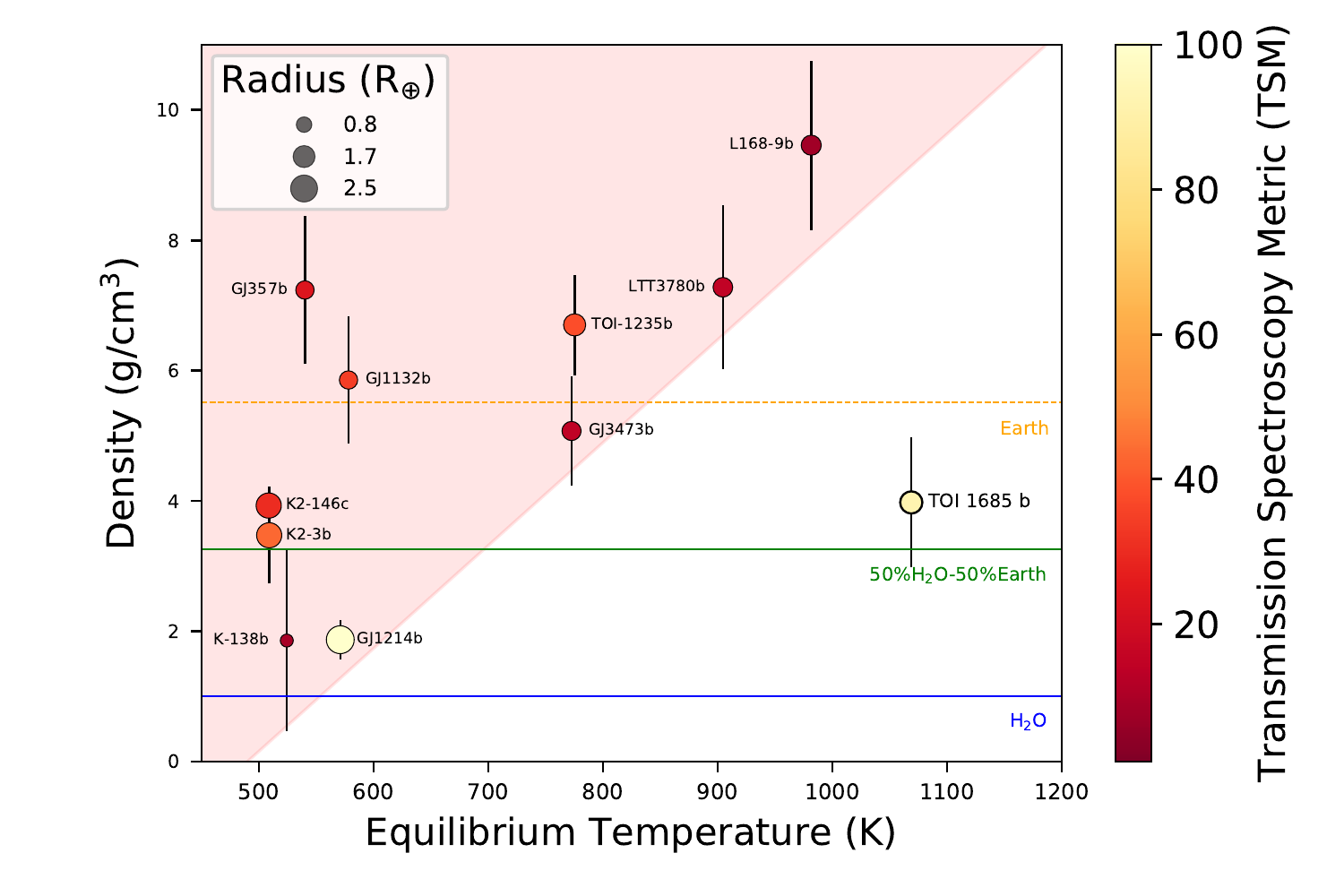}
      \caption{Density-equilibrium temperature diagram of warm-hot exoplanets with a radius of less than 3\,$R_\oplus$ around M dwarfs. All planets have mass determination by either RVs or TTVs. TOI-1685\,b is labeled and marked with a thicker black borderline. Some densities are marked as reference by horizontal lines, including Earth (dashed orange), H$_2$O (blue), and 50\,\% H$_2$O, 50\% silicate (green). The observed $\rho$-$T\rm_{eq}$ space before the discovery of TOI-1685\,b is shaded in red to illustrate the dissimilarity of TOI-1685\,b with  previous discoveries.}
         \label{fig:Density_Teq}
\end{figure}

Following such a scenario, it is expected that small exoplanets at higher temperatures have higher bulk densities. This is indeed what has been observed so far, as shown by the red shaded region in Fig.~\ref{fig:Density_Teq}. However, TOI-1685\,b does not follow such a prediction. This may suggest that TOI-1685\,b maintains a substantial atmosphere, unlike other hot small planets. In such a scenario, water, carbon dioxide, or methane features could be observable in its atmosphere \citep{molaverdikhani2019colda,molaverdikhani2019coldb} or such atmospheric features might be obscured by clouds \citep{molaverdikhani2020role}. Nevertheless, future emission and transmission spectroscopy of TOI-1685\,b is needed to answer the question of whether the entire atmosphere has escaped or a substantial atmosphere has been maintained on TOI-1685\,b, making this USP a rather unusual and interesting planet discovery.

\subsection{The planet candidate TOI-1685~[c]}
\label{subsec:Planet_c}

Our RV modeling shows moderate evidence for a second potentially planetary signal in the system. 
As  discussed in Sect,~\ref{subsec:Prot}, the observed period of $\sim9$\,d is not obviously linked to the stellar rotation period of $\sim$26\,d, and the analysis of a comprehensive set of activity indicators revealed no signs of stellar activity at the period in question. However, the fact that the $\sim$9\,d signal is close to the first harmonic of the likely rotational period derived from the RVs implies that we cannot be completely certain about its origin.

A comparison of the log-evidence values of the different models considered does not settle the issue as the differences between them are not highly significant. The situation is further complicated by the fact that sinusoidal or Keplerian models may also represent stellar activity well, especially over a relatively short time span. 

Nevertheless, the significance and the coherence of the 9\,d signal compared to the presumed stellar activity signal at 19\,d does lend some support to a planetary origin. After all, it represents the highest peak in the RV periodogram (Fig.~\ref{fig:GLS_CARMENES}), and it seems to be the most persistent (Fig.~\ref{fig: sBGLS_TOI1685}). Under the assumption that the signal is indeed due to a planet, we performed a 2cp+GP model fit. We used the same distribution prior as that presented in Table~\ref{tab:priors}, and for the signal at $\sim$~9\,d we used a uniform distribution between 8.5 and 9.5\,d.

From our joint fit, we derived for the planet candidate a period of $P_{\rm c}$~=~$9.03^{+ 0.10}_{- 0.12}$\,d and a minimum mass of $M_{\rm c}$~=~$9.2^{+ 2.1}_{- 2.1}$\,$M_{\oplus}$; additional planet parameters are reported in Table~\ref{tab:derivedparams}. The obtained parameters with the 2cp+GP model were consistent within one sigma with those derived from the 1cp+GP model. 

As a further test, we investigated whether the two-planet system would be dynamically stable. We used \texttt{Exo-Striker}\footnote{\url{https://github.com/3fon3fonov/exostriker}} \citep{2019ascl.soft06004T} to check the long-term stability of planetary systems via the angular momentum deficit key parameter \citep{2017A&A...605A..72L}. 
The best joint fit resulted in a stable solution for the TOI-1685 system.

\section{Conclusion} \label{sec:conclusion}

We present the discovery of a possible multi-planetary system around the M3.0\,V star TOI-1685. The system has one transiting planet with an ultra-short orbital period plus another planet candidate at a wider orbit found only in RV data. 
The USP TOI-1685\,b was first detected in the photometric time series of sector 19 of the {\it TESS} mission. We collected CARMENES RV data, as well as photometric transit follow-up observations from LCOGT, PESTO, and MuSCAT2, with which we confirmed its planetary nature. From the joint analysis, we derived a mass of $M_{\rm b}$~=~$3.78 ^{+ 0.63}_{- 0.63}$\,$M_\oplus$ and a radius of $R_{\rm b}$~=~$1.70^{+ 0.07}_{- 0.07}$\,$R_\oplus$. 
The derived  bulk density of $\rho_{\rm b}$~=~$4.21^{+0.95}_{-0.82}$\,g\,cm$^{-3}$ makes TOI-1685\,b the least dense USP around an M dwarf known to date.

A comparison of the physical properties of TOI-1685\,b with compositional models revealed a bulk composition of 50\,\% H$_{2}$0 and 50\,\% silicate. With a mass and radius precision better than 18\,\% and 5\,\%, respectively, TOI-1685\,b complements the sample of well-characterized small planets orbiting nearby M dwarfs. 
Its proximity to its host star and the measured values for the TSM and the ESM metrics qualify this planet for atmospheric characterization through emission and transmission spectroscopy, as well as make it an interesting planet for studying atmospheric evolution and escape processes.

In the exploration of the RV data, a significant signal at $\sim$9\,d was also found. To explore the origin of this signal, we analyzed the periodogram for RV activity indicators as well as the s-BGLS periodogram, and the signal was found to be persistent.

To model the stellar activity we used a GP-QP model based on a QP kernel plus two circular orbits (2cp+GP). However, due to the proximity of the $\sim$9\,d planet candidate period to half of the stellar rotation period, we cannot rule out that it is related to stellar activity. Nevertheless, the strength and coherence of the signal make it a promising planet candidate.
However, based on the currently available RV data, it is not possible to confidently claim the detection of a second planet in the system. To reach a solid conclusion, more data will be needed.

\begin{acknowledgements}

CARMENES is an instrument at the Centro Astron\'omico Hispano-Alem\'an (CAHA) at Calar Alto (Almer\'{\i}a, Spain), operated jointly by the Junta de Andaluc\'ia and the Instituto de Astrof\'isica de Andaluc\'ia (CSIC).
CARMENES was funded by the Max-Planck-Gesellschaft (MPG), 
  the Consejo Superior de Investigaciones Cient\'{\i}ficas (CSIC),
  the Ministerio de Econom\'ia y Competitividad (MINECO) and the European Regional Development Fund (ERDF) through projects FICTS-2011-02, ICTS-2017-07-CAHA-4, and CAHA16-CE-3978, 
  and the members of the CARMENES Consortium 
  (Max-Planck-Institut f\"ur Astronomie,
  Instituto de Astrof\'{\i}sica de Andaluc\'{\i}a,
  Landessternwarte K\"onigstuhl,
  Institut de Ci\`encies de l'Espai,
  Institut f\"ur Astrophysik G\"ottingen,
  Universidad Complutense de Madrid,
  Th\"uringer Landessternwarte Tautenburg,
  Instituto de Astrof\'{\i}sica de Canarias,
  Hamburger Sternwarte,
  Centro de Astrobiolog\'{\i}a and
  Centro Astron\'omico Hispano-Alem\'an), 
  with additional contributions by the MINECO, 
  the Deutsche Forschungsgemeinschaft through the Major Research Instrumentation Programme and Research Unit FOR2544 ``Blue Planets around Red Stars'', 
  the Klaus Tschira Stiftung, 
  the states of Baden-W\"urttemberg and Niedersachsen, 
  and by the Junta de Andaluc\'{\i}a.
  
We acknowledge financial support from the Agencia Estatal de Investigaci\'on of the Ministerio de Ciencia, Innovaci\'on y Universidades and the ERDF through projects 
  PID2019-109522GB-C5[1:4]/AEI/10.13039/501100011033,   
  PGC2018-098153-B-C33,                         
and the Centre of Excellence ``Severo Ochoa'' and ``Mar\'ia de Maeztu'' awards to the Instituto de Astrof\'isica de Canarias (SEV-2015-0548), Instituto de Astrof\'isica de Andaluc\'ia (SEV-2017-0709), and Centro de Astrobiolog\'ia (MDM-2017-0737), 
the Generalitat de Catalunya/CERCA programme,
``la Caixa'' Foundation (100010434),
European Union's Horizon 2020 research and innovation programme under the Marie Skłodowska-Curie grant agreement No. 847648 (LCF/BQ/PI20/11760023),
a University Research Support Grant from the National Astronomical Observatory of Japan,
JSPS KAKENHI (JP15H02063, JP18H01265, JP18H05439, JP18H05442, and JP22000005),
JST PRESTO (JPMJPR1775),
UK Science and Technology Facilities Council (ST/R000824/1),
and NASA (NNX17AG24G)

This article is based on observations made with the MuSCAT2 instrument, developed by ABC, at Telescopio Carlos S\'anchez operated on the island of Tenerife by the IAC in the Spanish Observatorio del Teide,
with the Observatoire du Mont-M\'egantic, financed by Universit\'e de Montr\'eal, Universit\'e Laval, the National Sciences and Engineering Council of Canada, the Fonds qu\'eb\'ecois de la recherche sur la Nature et les technologies, and the Canada Economic Development program,
and with the LCOGT network.
LCOGT telescope time was granted by NOIRLab through the Mid-Scale Innovations Program , which is funded by NSF.

Funding for the TESS mission is provided by NASA's Science Mission Directorate.  
We acknowledge the use of public TESS data from pipelines at the TESS Science Office and at the TESS Science Processing Operations Center.
This research has made use of the Exoplanet Follow-up Observation Program website, which is operated by the California Institute of Technology, under contract with the National Aeronautics and Space Administration under the Exoplanet Exploration Program.

This research also made use of \texttt{AstroImageJ}, and \texttt{TAPIR}, and the SIMBAD database, operated at CDS, Strasbourg, France.

We thank the referee for the careful report and Carlos Cifuentes for computing $L_{\rm bol}$.
Special thanks to Ismael Pessa
for all his support through this work and Mat\'ias Jones for being always available to answer questions from an old student and friend.
\end{acknowledgements}

\bibliographystyle{aa} 
\bibliography{biblio} 

\begin{appendix}

\section{Additional figures and tables}

\begin{table*}
\centering
\small
\caption{RV measurements and spectroscopic activity indicators for TOI-1685 from CARMENES VIS spectra.} 
\label{tab:RV_Activity_all}
\begin{tabular}{c c c c c c c c c}
\hline
\hline
        \noalign{\smallskip}
        BJD & RV & CRX & dLW & H$\alpha$ & Ca~IRTa & TiO7050 & TiO8430 & TiO8860 \\
        (--2450000) & (m\,s$^{-1}$) & (m\,s$^{-1}$\,Np$^{-1}$) & (m$^{2}$\,s$^{-2}$) &\\
        \noalign{\smallskip}
        \hline
        \noalign{\smallskip}
9069.6744 & -1.87$\pm$2.43 & 9.17$\pm$24.55 & 23.23$\pm$3.44 & 0.8715$\pm$0.0043 & 0.6121$\pm$0.0031 & 0.616$\pm$0.002 & 0.836$\pm$0.004 & 0.004$\pm$0.003 \\
9073.6703 & 8.05$\pm$4.91 & 30.68$\pm$50.73 & 23.22$\pm$5.48 & 0.8531$\pm$0.0095 & 0.6132$\pm$0.0067 & 0.620$\pm$0.004 & 0.822$\pm$0.008 & 0.008$\pm$0.006 \\
9095.6713 & -2.74$\pm$2.14 & -3.32$\pm$19.66 & 4.05$\pm$2.91 & 0.8605$\pm$0.0037 & 0.6124$\pm$0.0029 & 0.616$\pm$0.002 & 0.831$\pm$0.003 & 0.003$\pm$0.003 \\
9097.6749 & 14.20$\pm$2.81 & 12.22$\pm$28.42 & -7.50$\pm$3.14 & 0.8794$\pm$0.0059 & 0.6111$\pm$0.0044 & 0.612$\pm$0.003 & 0.834$\pm$0.005 & 0.005$\pm$0.004 \\
9098.6754 & 5.97$\pm$1.78 & 21.89$\pm$14.47 & -1.50$\pm$2.17 & 0.8591$\pm$0.0029 & 0.6054$\pm$0.0023 & 0.619$\pm$0.001 & 0.838$\pm$0.003 & 0.003$\pm$0.002 \\
9099.6693 & -1.93$\pm$2.13 & -23.05$\pm$20.72 & 0.91$\pm$2.25 & 0.8480$\pm$0.0031 & 0.6069$\pm$0.0025 & 0.616$\pm$0.001 & 0.840$\pm$0.003 & 0.003$\pm$0.002 \\
9101.6859 & -0.34$\pm$2.63 & -20.53$\pm$24.23 & 1.27$\pm$3.41 & 0.8539$\pm$0.0050 & 0.6010$\pm$0.0037 & 0.616$\pm$0.002 & 0.835$\pm$0.004 & 0.004$\pm$0.004 \\
9102.6845 & -15.14$\pm$3.71 & 45.34$\pm$26.79 & 0.25$\pm$2.40 & 0.8520$\pm$0.0047 & 0.6002$\pm$0.0035 & 0.616$\pm$0.002 & 0.829$\pm$0.004 & 0.004$\pm$0.004 \\
9103.6777 & 1.44$\pm$1.84 & 11.14$\pm$14.56 & -2.91$\pm$1.79 & 0.8635$\pm$0.0030 & 0.6062$\pm$0.0024 & 0.619$\pm$0.001 & 0.840$\pm$0.003 & 0.003$\pm$0.002 \\
9114.7106 & 10.09$\pm$3.26 & -20.69$\pm$33.19 & -10.15$\pm$3.45 & 0.8709$\pm$0.0058 & 0.6069$\pm$0.0042 & 0.618$\pm$0.002 & 0.841$\pm$0.005 & 0.005$\pm$0.004 \\
9118.6966 & 0.46$\pm$2.46 & 83.69$\pm$15.50 & 0.61$\pm$2.41 & 0.8573$\pm$0.0040 & 0.6134$\pm$0.0031 & 0.617$\pm$0.002 & 0.832$\pm$0.003 & 0.003$\pm$0.003 \\
9120.6746 & -7.34$\pm$1.95 & 32.41$\pm$17.54 & -4.17$\pm$1.82 & 0.8422$\pm$0.0030 & 0.6135$\pm$0.0024 & 0.617$\pm$0.001 & 0.834$\pm$0.003 & 0.003$\pm$0.002 \\
9121.6352 & -10.46$\pm$1.90 & -24.59$\pm$17.06 & -0.05$\pm$1.94 & 0.8777$\pm$0.0033 & 0.6192$\pm$0.0025 & 0.618$\pm$0.001 & 0.833$\pm$0.003 & 0.003$\pm$0.003 \\
9122.6841 & -0.98$\pm$1.82 & 25.19$\pm$17.66 & -1.04$\pm$2.68 & 0.8602$\pm$0.0046 & 0.6126$\pm$0.0035 & 0.617$\pm$0.002 & 0.838$\pm$0.004 & 0.004$\pm$0.003 \\
9127.6877 & -1.46$\pm$2.61 & -28.39$\pm$22.25 & 0.54$\pm$2.89 & 0.8678$\pm$0.0041 & 0.6235$\pm$0.0032 & 0.616$\pm$0.002 & 0.835$\pm$0.004 & 0.004$\pm$0.003 \\
9128.6223 & 2.87$\pm$1.83 & -41.62$\pm$14.62 & 1.58$\pm$1.89 & 0.8685$\pm$0.0033 & 0.6157$\pm$0.0026 & 0.617$\pm$0.001 & 0.836$\pm$0.003 & 0.003$\pm$0.003 \\
9131.6723 & 1.86$\pm$1.81 & -15.46$\pm$16.06 & -0.81$\pm$1.89 & 0.8874$\pm$0.0031 & 0.6121$\pm$0.0024 & 0.616$\pm$0.001 & 0.837$\pm$0.003 & 0.003$\pm$0.002 \\
9132.6702 & 10.82$\pm$2.8 & 7.56$\pm$28.28 & 4.81$\pm$3.04 & 0.9270$\pm$0.0050 & 0.6233$\pm$0.0037 & 0.617$\pm$0.002 & 0.838$\pm$0.004 & 0.004$\pm$0.004 \\
9138.6485 & 5.90$\pm$3.33 & -21.62$\pm$34.82 & -10.22$\pm$4.55 & 0.9023$\pm$0.0079 & 0.6240$\pm$0.0058 & 0.613$\pm$0.003 & 0.832$\pm$0.006 & 0.006$\pm$0.006 \\
9139.4464 & -5.21$\pm$2.91 & 62.16$\pm$29.17 & -9.47$\pm$3.24 & 0.9378$\pm$0.0053 & 0.6143$\pm$0.0038 & 0.623$\pm$0.002 & 0.827$\pm$0.004 & 0.004$\pm$0.004 \\
9139.5469 & -7.91$\pm$1.66 & -13.62$\pm$16.14 & -3.51$\pm$2.18 & 0.8794$\pm$0.0038 & 0.6089$\pm$0.0029 & 0.614$\pm$0.002 & 0.835$\pm$0.003 & 0.003$\pm$0.003 \\
9139.6241 & -4.77$\pm$2.93 & -34.23$\pm$29.98 & 2.84$\pm$3.54 & 0.8765$\pm$0.0047 & 0.6088$\pm$0.0035 & 0.610$\pm$0.002 & 0.826$\pm$0.004 & 0.004$\pm$0.003 \\
9139.7292 & -5.76$\pm$1.91 & 16.39$\pm$18.42 & 20.24$\pm$3.11 & 0.8773$\pm$0.0038 & 0.6246$\pm$0.0032 & 0.620$\pm$0.002 & 0.836$\pm$0.004 & 0.004$\pm$0.003 \\
9140.5196 & -4.64$\pm$1.54 & -13.22$\pm$12.95 & 0.0$\pm$1.71 & 0.8837$\pm$0.0029 & 0.6159$\pm$0.0023 & 0.612$\pm$0.001 & 0.829$\pm$0.002 & 0.002$\pm$0.002 \\
9140.5965 & -3.62$\pm$1.65 & -0.46$\pm$14.53 & -6.21$\pm$1.79 & 0.8840$\pm$0.0029 & 0.6201$\pm$0.0023 & 0.613$\pm$0.001 & 0.832$\pm$0.003 & 0.003$\pm$0.002 \\
9140.6963 & -3.50$\pm$1.72 & 16.93$\pm$14.90 & -4.29$\pm$1.97 & 0.9025$\pm$0.0034 & 0.6230$\pm$0.0027 & 0.616$\pm$0.002 & 0.825$\pm$0.003 & 0.003$\pm$0.003 \\
9141.5171 & -11.99$\pm$2.8 & 20.14$\pm$25.05 & -8.89$\pm$2.51 & 0.8737$\pm$0.0042 & 0.6242$\pm$0.0032 & 0.613$\pm$0.002 & 0.828$\pm$0.003 & 0.003$\pm$0.003 \\
9141.5792 & -9.55$\pm$2.78 & -30.48$\pm$24.65 & -7.91$\pm$2.73 & 0.8732$\pm$0.0048 & 0.6133$\pm$0.0036 & 0.618$\pm$0.002 & 0.831$\pm$0.004 & 0.004$\pm$0.004 \\
9141.6397 & -7.51$\pm$3.00 & 47.54$\pm$26.84 & -7.12$\pm$2.60 & 0.8755$\pm$0.0047 & 0.6164$\pm$0.0035 & 0.618$\pm$0.002 & 0.832$\pm$0.004 & 0.004$\pm$0.004 \\
9141.7027 & -8.45$\pm$2.59 & -1.09$\pm$22.58 & -9.26$\pm$2.69 & 0.8756$\pm$0.0044 & 0.6143$\pm$0.0037 & 0.615$\pm$0.002 & 0.837$\pm$0.004 & 0.004$\pm$0.004 \\
9142.5187 & 7.29$\pm$2.74 & 11.59$\pm$27.32 & 0.14$\pm$3.83 & 0.8788$\pm$0.0055 & 0.6051$\pm$0.0039 & 0.611$\pm$0.002 & 0.844$\pm$0.005 & 0.005$\pm$0.004 \\
9146.5184 & -0.12$\pm$2.91 & 24.29$\pm$30.27 & -9.94$\pm$3.42 & 0.8472$\pm$0.0060 & 0.5986$\pm$0.0044 & 0.622$\pm$0.003 & 0.815$\pm$0.005 & 0.005$\pm$0.004 \\
9146.6025 & 7.18$\pm$1.98 & -5.05$\pm$19.81 & -2.67$\pm$3.28 & 0.8671$\pm$0.0051 & 0.6118$\pm$0.0039 & 0.620$\pm$0.002 & 0.831$\pm$0.004 & 0.004$\pm$0.004 \\
9147.4080 & -3.36$\pm$2.12 & 11.37$\pm$21.39 & 12.64$\pm$3.18 & 0.8812$\pm$0.0058 & 0.6122$\pm$0.0041 & 0.621$\pm$0.002 & 0.816$\pm$0.005 & 0.005$\pm$0.004 \\
9147.5126 & -5.02$\pm$2.86 & 12.22$\pm$29.49 & -3.44$\pm$3.67 & 0.8605$\pm$0.0062 & 0.6149$\pm$0.0043 & 0.614$\pm$0.003 & 0.827$\pm$0.005 & 0.005$\pm$0.004 \\
9149.4108 & -3.6$\pm$2.95 & 41.15$\pm$30.27 & -2.14$\pm$3.11 & 0.8765$\pm$0.0054 & 0.6174$\pm$0.0039 & 0.618$\pm$0.002 & 0.834$\pm$0.003 & 0.003$\pm$0.003 \\
9149.5024 & -5.31$\pm$2.48 & -15.23$\pm$24.06 & 7.44$\pm$3.84 & 0.8563$\pm$0.0048 & 0.6117$\pm$0.0036 & 0.621$\pm$0.002 & 0.822$\pm$0.004 & 0.004$\pm$0.004 \\
9149.5915 & -6.27$\pm$2.76 & 10.11$\pm$28.16 & -1.92$\pm$3.67 & 0.8749$\pm$0.0051 & 0.6117$\pm$0.0039 & 0.629$\pm$0.002 & 0.841$\pm$0.004 & 0.004$\pm$0.004 \\
9149.6962 & -3.53$\pm$3.95 & -20.06$\pm$41.52 & -38.04$\pm$5.54 & 0.8493$\pm$0.0073 & 0.6071$\pm$0.0058 & 0.616$\pm$0.002 & 0.833$\pm$0.004 & 0.004$\pm$0.004 \\
9150.3895 & -5.06$\pm$2.30 & -16.21$\pm$23.57 & 3.44$\pm$2.26 & 0.8578$\pm$0.0041 & 0.6025$\pm$0.0032 & 0.610$\pm$0.003 & 0.839$\pm$0.007 & 0.007$\pm$0.006 \\
9151.6239 & 0.51$\pm$1.78 & 3.03$\pm$16.67 & -2.03$\pm$2.00 & 0.8627$\pm$0.0033 & 0.6118$\pm$0.0026 & 0.611$\pm$0.002 & 0.830$\pm$0.004 & 0.004$\pm$0.003 \\
9151.7309 & 10.76$\pm$1.97 & -26.45$\pm$18.18 & -2.60$\pm$2.27 & 0.8928$\pm$0.0040 & 0.6265$\pm$0.0032 & 0.617$\pm$0.001 & 0.830$\pm$0.003 & 0.003$\pm$0.003 \\
9152.4645 & 7.73$\pm$1.74 & -10.62$\pm$16.92 & 1.94$\pm$1.92 & 0.8654$\pm$0.0029 & 0.6183$\pm$0.0023 & 0.619$\pm$0.002 & 0.837$\pm$0.004 & 0.004$\pm$0.003 \\
9153.3820 & 9.91$\pm$2.40 & -26.93$\pm$23.87 & 10.93$\pm$2.72 & 0.8557$\pm$0.0037 & 0.6090$\pm$0.0029 & 0.617$\pm$0.001 & 0.832$\pm$0.003 & 0.003$\pm$0.002 \\
9153.4717 & 9.87$\pm$2.05 & -13.32$\pm$18.68 & 20.78$\pm$1.97 & 0.9707$\pm$0.0032 & 0.6261$\pm$0.0024 & 0.616$\pm$0.002 & 0.833$\pm$0.003 & 0.003$\pm$0.003 \\
9154.4992 & 4.88$\pm$1.76 & 5.60$\pm$17.37 & 8.33$\pm$1.60 & 0.8606$\pm$0.0029 & 0.6142$\pm$0.0023 & 0.618$\pm$0.001 & 0.838$\pm$0.003 & 0.003$\pm$0.002 \\
9154.6201 & 7.08$\pm$1.73 & 7.89$\pm$16.85 & 11.88$\pm$2.64 & 0.8660$\pm$0.0034 & 0.6192$\pm$0.0028 & 0.614$\pm$0.001 & 0.83$\pm$0.003 & 0.003$\pm$0.002 \\
9156.4518 & 1.17$\pm$1.60 & -10.93$\pm$14.04 & 1.14$\pm$2.04 & 0.8961$\pm$0.0038 & 0.6187$\pm$0.0030 & 0.617$\pm$0.002 & 0.833$\pm$0.003 & 0.003$\pm$0.003 \\
9156.5853 & 3.51$\pm$1.61 & 8.04$\pm$14.35 & 3.97$\pm$1.66 & 0.8785$\pm$0.0027 & 0.6094$\pm$0.0022 & 0.615$\pm$0.002 & 0.822$\pm$0.003 & 0.003$\pm$0.003 \\
9161.3599 & 8.37$\pm$3.26 & -29.20$\pm$35.50 & -47.90$\pm$7.13 & 0.8561$\pm$0.0067 & 0.6055$\pm$0.0047 & 0.615$\pm$0.001 & 0.833$\pm$0.002 & 0.002$\pm$0.002 \\
9161.4505 & 9.47$\pm$3.97 & 34.02$\pm$42.22 & -12.87$\pm$4.06 & 0.8620$\pm$0.0076 & 0.6142$\pm$0.0055 & .... & ..... & ..... \\
9161.5732 & 0.63$\pm$1.63 & 13.21$\pm$15.68 & -5.77$\pm$2.58 & 0.8814$\pm$0.0042 & 0.6113$\pm$0.0032 & .... & .... & ..... \\
9161.6724 & -4.24$\pm$2.18 & -25.79$\pm$21.32 & -8.84$\pm$2.16 & 0.8623$\pm$0.0039 & 0.6153$\pm$0.0032 & .... & ..... & .... \\
9163.3774 & 5.92$\pm$2.99 & -42.62$\pm$31.26 & -8.72$\pm$3.81 & 0.9328$\pm$0.0069 & 0.6167$\pm$0.0049 & .... & ..... & .... \\
9163.4992 & -1.73$\pm$3.03 & -76.62$\pm$29.37 & -14.60$\pm$3.82 & 0.8851$\pm$0.0062 & 0.6183$\pm$0.0045 & .... & ..... & .... \\

\hline
\end{tabular}
\end{table*}

\begin{table*}
    \centering
    \caption{Priors used for TOI-1685\,b and TOI-1685\,[c] in the  joint fit with \texttt{juliet}.}
    \label{tab:priors}
    \begin{tabular}{lccr} 
        \hline
        \hline
        \noalign{\smallskip}
        Parameter$^a$  & Prior & Unit & Description \\
        \noalign{\smallskip}
        \hline
        \noalign{\smallskip}
        \multicolumn{4}{c}{\it Stellar parameters} \\
        \noalign{\smallskip}
        $\rho_\star$ & $\mathcal{U}(5.7,5.9)$ & g\,cm\,$^{-3}$ & Stellar density \\
        \noalign{\smallskip}
        \multicolumn{4}{c}{\it Planet b parameters} \\
        \noalign{\smallskip}
        $P_{\rm b}$              & $\mathcal{N}(0.66,0.01)$           & d                    & Period of planet b \\
        $t_{0,b}$      & $\mathcal{U}(2458816.0,2458816.7)$     & d                    & Time of transit center of planet b \\
        $r_{1,b}$                & $\mathcal{U}(0,1)$           & \dots                & Parameterization for $p$ and $b$ \\
        $r_{2,b}$                & $\mathcal{U}(0,1)$           & \dots                & Parameterization for $p$ and $b$ \\
        $K_{b}$                  & $\mathcal{U}(0,10)$        & $\mathrm{m\,s^{-1}}$ & RV semi-amplitude of planet b \\
        $e_{b}$                & 0.0 (fixed)                    & \dots                & Orbital eccentricity of planet b \\
        $\omega_{b}$              & 90.0 (fixed)                    & deg              & Periastron angle of planet b \\
        \noalign{\smallskip}
        \multicolumn{4}{c}{\it Planet candidate [c] parameters --> only used for 2cp+GP model fit} \\
        \noalign{\smallskip}        
        $P_{\rm c}$              & $\mathcal{U}(8.5,9.5)$           & d                    & Period of candidate [c] \\
        $t_{0,c}$      & $\mathcal{U}(2458816,2458826)$     & d                    & Time of transit center of candidate [c] \\
        $K_{c}$                  & $\mathcal{U}(0,10)$        & $\mathrm{m\,s^{-1}}$ & RV semi-amplitude of candidate [c] \\
        $e_{c}$                & 0.0 (fixed)                    & \dots                & Orbital eccentricity of candidate [c] \\
        $\omega_{c}$              & 90.0 (fixed)                    & deg              & Periastron angle of candidate [c] \\
        \noalign{\smallskip}
        \multicolumn{4}{c}{\it Photometry parameters for TESS Sector 19} \\
        \noalign{\smallskip}
        $D_{\mathrm{TESS}}$         & 1.0 (fixed)             & \dots     & Dilution factor for {\em TESS} \\
        $M_{\mathrm{TESS}}$                & $\mathcal{N}(0,0.1)$    & \dots       & Relative flux offset for {\em TESS} \\
        $\sigma_{\mathrm{TESS}}$           & $\mathcal{J}(0.1,1000)$ & ppm       & Extra jitter term for {\em TESS} \\
        $q_{1,\mathrm{TESS}}$        & $\mathcal{U}(0,1)$      & \dots     & Limb-darkening parameterization for {\em TESS} \\
        $q_{2,\mathrm{TESS}}$        & $\mathcal{U}(0,1)$      & \dots     & Limb-darkening parameterization for {\em TESS} \\
        \noalign{\smallskip}
        \multicolumn{4}{c}{\it Photometry parameters for LCOGT nights, 2020-08-26, 2020-11-07 and, 2020-11-11} \\
        \noalign{\smallskip}        
        $D_{\mathrm{LCOGT}}$                   & 1.0 (fixed)             & \dots     & Dilution factor for LCOGT\\
        $M_{\mathrm{LCOGT}}$                   & $\mathcal{N}(0,0.1)$    & \dots       & Relative flux offset for LCOGT\\
        $\sigma_{\mathrm{LCOGT}}$              & $\mathcal{J}(0.1,1000)$ & ppm       & Extra jitter term for LCOGT\\
        $q_{1,\mathrm{LCOGT         }}$        & $\mathcal{U}(0,1)$      & \dots     & Limb-darkening parameterization for LCOGT\\ 
        \noalign{\smallskip}
        \multicolumn{4}{c}{\it Photometry parameters for PESTO, night 2020-03-08} \\
        \noalign{\smallskip}        
        $D_{\mathrm{PESTO}}$                   & 1.0 (fixed)             & \dots     & Dilution factor for PESTO\\
        $M_{\mathrm{PESTO}}$                   & $\mathcal{N}(0,0.1)$    & \dots       & Relative flux offset for PESTO\\
        $\sigma_{\mathrm{PESTO}}$              & $\mathcal{J}(0.1,1000)$ & ppm       & Extra jitter term for PESTO\\
        $q_{1,\mathrm{PESTO}}$        & $\mathcal{U}(0,1)$      & \dots     & Limb-darkening parameterization for PESTO\\ 
        \noalign{\smallskip}
        \multicolumn{4}{c}{\it Photometry parameters for MuSCATS2 i, and z$_s$ bands, night 2021-01-19 and 2021-01-29 } \\
        \noalign{\smallskip}      
        $D_{\mathrm{MuSCAT2~(i,~z_s)}}$                   & 1.0 (fixed)             & \dots     & Dilution factor for MuSCAT2\\
        $M_{\mathrm{MuSCAT2~(i,~z_s)}}$                   & $\mathcal{N}(0,0.1)$    & \dots       & Relative flux offset for MuSCAT2\\
        $\sigma_{\mathrm{MuSCAT2~(i,~z_s)}}$              & $\mathcal{J}(0.1,1000)$ & ppm       & Extra jitter term for MuSCAT2\\
        $q_{1,\mathrm{LCOGT~(i,~z_s)}}$        & $\mathcal{U}(0,1)$      & \dots     & Limb-darkening parameterization\\         
        \noalign{\smallskip}
        \multicolumn{4}{c}{\it RV parameters} \\
        \noalign{\smallskip}
        $\gamma_{\mathrm{CARMENES}}$            & $\mathcal{U}(-10,10)$     & $\mathrm{m\,s^{-1}}$ & RV zero point for CARMENES \\
        $\sigma_{\mathrm{CARMENES}}$         & $\mathcal{J}(0.01,10)$ & $\mathrm{m\,s^{-1}}$ & Extra jitter term for CARMENES \\
        \noalign{\smallskip}
        \noalign{\smallskip}
        \multicolumn{4}{c}{\it GP hyperparameters} \\
        \noalign{\smallskip}
        $\sigma_\mathrm{GP,RV}$                  & $\mathcal{U}(0,80)$  & $\mathrm{m\,s^{-1}}$  & Amplitude of GP component for the RVs \\
        $\alpha_\mathrm{GP,RV}$                  & $\mathcal{J}(10^{-10},0.01)$  & d$^{-2}$              & Inverse length scale of GP exponential component for the RVs \\
        $\Gamma_\mathrm{GP,RV}$                  & $\mathcal{J}(0.1,10)$  & \dots                  & Amplitude of GP sine-squared component for the RVs \\
        $P_\mathrm{rot;GP,RV}$                   & $\mathcal{U}(15,30)$          & d                     & Period of the GP quasi-periodic component for the RVs \\
        \noalign{\smallskip}        
        \hline
    \end{tabular}
     \tablefoot{
        \tablefoottext{a}{The parameterization for $(p,b)$ was made with $(r_1,r_2)$ as in \citet{Espinoza18}. 
        The prior labels of $\mathcal{N}$, $\mathcal{U}$, and $\mathcal{J}$ represent normal, uniform, and Jeffreys (log-uniform) distributions, respectively;
        $\mathcal{N}(\mu,\sigma^2)$ is a normal distribution of the mean, $\mu$, and variance, $\sigma^2$; $\mathcal{U}(a,b)$ and $\mathcal{J}(a,b)$ are uniform; and Jeffreys (log-uniform) distributions are between $a$ and $b$}.}
\end{table*}

\begin{figure*}
    \centering
    \includegraphics[width=\textwidth]{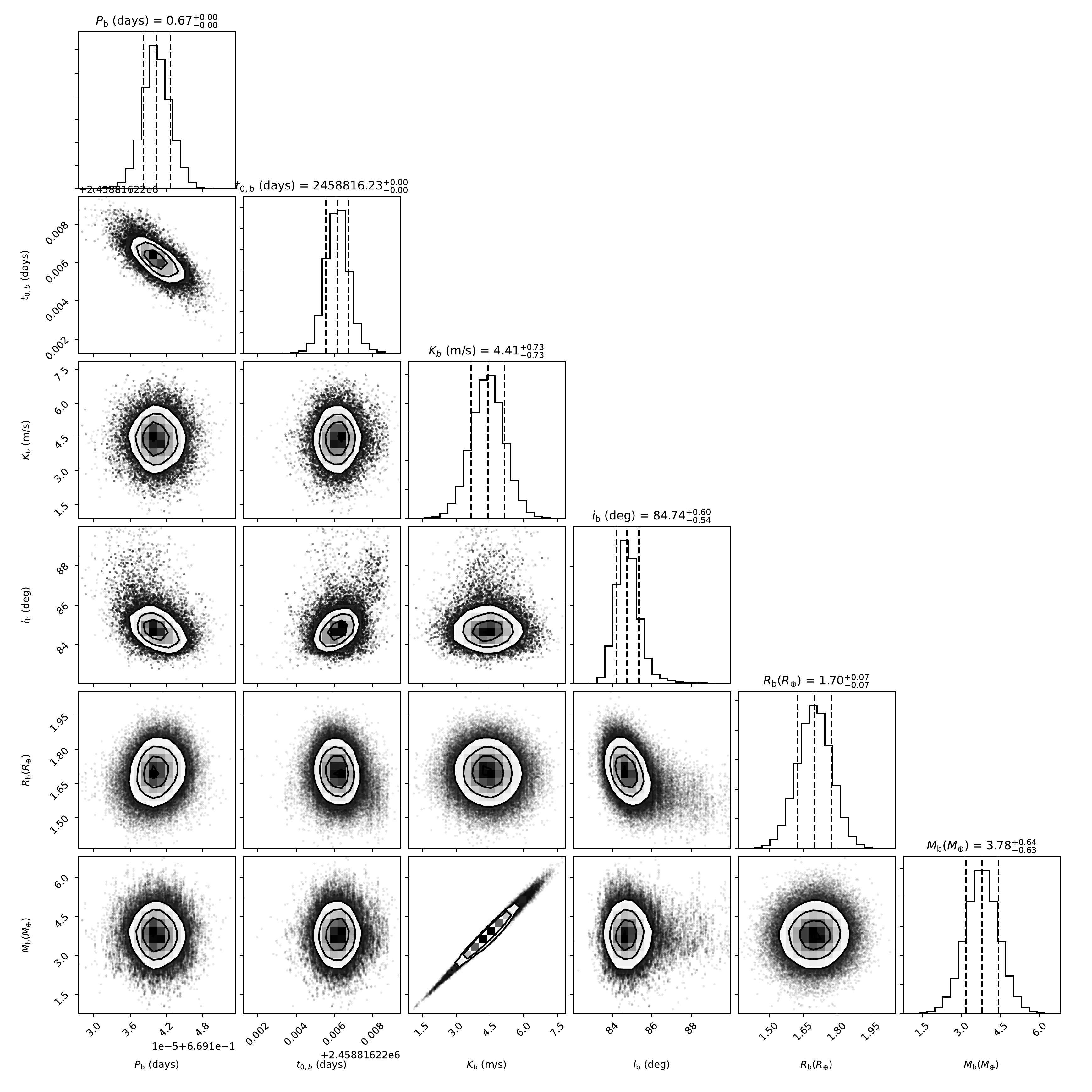}
    \caption{Posterior distribution for the joint model parameters (1cp+GP) derived with \texttt{juliet}.}
    \label{fig:corner_plot}
\end{figure*}

\end{appendix}

\end{document}